\documentclass[utf8]{IEEEtran}

\usepackage{xspace}
\usepackage[bookmarks=false]{hyperref}
\usepackage{booktabs}		% \toprule, \midrule, \bottomrule
\usepackage[utf8]{inputenc}   % switch your editors to UTF-8 for umlauts (ck)
\usepackage{amsmath}
\usepackage{amssymb}
\usepackage{xcolor}
\usepackage{graphicx}
\usepackage{subcaption} % subfigures
\usepackage{colortbl}
\usepackage{multirow}
\usepackage{amsthm}
\usepackage[pagewise,displaymath, mathlines]{lineno}

\newcommand*\patchAmsMathEnvironmentForLineno[1]{%
  \expandafter\let\csname old#1\expandafter\endcsname\csname #1\endcsname
  \expandafter\let\csname oldend#1\expandafter\endcsname\csname end#1\endcsname
  \renewenvironment{#1}%
     {\linenomath\csname old#1\endcsname}%
     {\csname oldend#1\endcsname\endlinenomath}}% 
\definecolor{mygreen}{rgb}{0.0745,0.6314,0.1177}
\definecolor{myblue}{rgb}{0.0745,0.0745,0.6314}
\definecolor{myred}{rgb}{0.62745,0,0}
\definecolor{darkgrey}{HTML}{888888}
\definecolor{grey}{HTML}{BBBBBB}
\definecolor{lightgrey}{HTML}{F7F7F7}
\definecolor{himmel}{HTML}{E6E6E6}
\definecolor{randtextcolor}{HTML}{FFFFFF}
\definecolor{blau}{HTML}{33FFFF}

\definecolor{mydodgerblue}{HTML}{1E90FF}
\definecolor{myindianred}{HTML}{CD5C5C}
\definecolor{myMtencolor}{HTML}{00BFFF}

\newcommand{\tblue}[1]{\textcolor{myblue}{#1}}
\newcommand{\tred}[1]{\textcolor{myred}{#1}}
\newcommand{\tgreen}[1]{\textcolor{mygreen}{#1}}
\renewcommand{\tblue}[1]{#1}
\renewcommand{\tred}[1]{#1}
\renewcommand{\tgreen}[1]{#1}

\newcommand{\rowGrey}{\rowcolor[gray]{0.85}}

\newcommand{\mxpr}{x_{\text{prol}}}
\newcommand{\mxtr}{x_{\text{tr}}}

\newcommand{\mxma}{x_{\text{wbc}}}

\newcommand{\mxG}{x_{\text{g}}}
\newcommand{\mkin}{k_{e,g} + k_{\text{ANC}}\;\mbase}% {k_{\text{in}}}
\newcommand{\mkoutLeukemic}{k_{e,g} + k_{\text{ANC}}\;(x_{l2}+x_{wbc} )}%  {k_{\text{out}}}
\newcommand{\mbaseg}{B_{\text{g}}}
\newcommand{\feedbackBeta}{\left(\frac{\mxG}{\mbaseg}\right)^{\mbeta}}

\newcommand{\feedbackGcsfCancer}{\frac{1}{1+ c_1 \; \mxma + c_2 \; x_{l2}} }
\newcommand{\feedbackGamma}{\left(\frac{\mxG}{\mbaseg}\right)^{\mgamma}}

\newcommand{\mktr}{\tgreen{k_\text{tr}}}
\newcommand{\mbase}{\tgreen{B}}

\newcommand{\mgamma}{\tgreen{\gamma}}
\newcommand{\mbeta}{\tgreen{\beta}}
\newcommand{\mslope}{\tgreen{\text{slope}}}
\newcommand{\mslopeGamma}{\text{slopeG}}

\newcommand{\mkpr}{k_\text{prol}}

\newcommand{\muExpr}{\frac{\tblue{u_{c}} \; \mBSA}{\mduration}}

\newcommand{\mkma}{\tred{k_\text{wbc}}}
\newcommand{\mkaz}{\tred{k_{10}}}
\newcommand{\mkab}{\tred{k_{12}}}
\newcommand{\mkba}{\tred{k_{21}}}

\newcommand{\mV}{\tred{V_c}}
\newcommand{\mMM}{\tred{MM_\text{AraC}}}
\newcommand{\mBSA}{\tred{\text{BSA}}}
\newcommand{\mduration}{\text{dur}_c}
\newcommand{\mdosage}{\tred{\text{dosage}}}

\newcommand{\specnature}[1]{\noindent \textcolor{blue}{Specifications by Nature: #1}\medskip }
\renewcommand{\specnature}[1]{}

\newcommand{\squishlist}{
   \begin{list}{$\bullet$}
    { \setlength{\itemsep}{0pt}      \setlength{\parsep}{0pt}
      \setlength{\topsep}{0pt}       \setlength{\partopsep}{0pt}
      \setlength{\leftmargin}{1.5em} \setlength{\labelwidth}{1em}
      \setlength{\labelsep}{0.5em} } }
\newcommand{\squishend}{
    \end{list}  }

\newcommand{\bg}{\begin {eqnarray*}}
\newcommand{\eg}{\end {eqnarray*}}

\usepackage{tikz}
\usetikzlibrary{intersections}
\usepackage{pgfplots,pgfplotstable}

\usetikzlibrary{arrows,positioning,decorations.markings,decorations.pathreplacing,shapes}
\usetikzlibrary{3d,calc}
\usetikzlibrary{patterns}
\tikzstyle{vecArrow} = [thick, decoration=
	{markings,mark=at position 1 with
		{\arrow[semithick]{open triangle 60}}},double distance=1.4pt, shorten >= 5.5pt,
		preaction = {decorate},
		postaction = {draw,line width=1.4pt, white,shorten >= 4.5pt}]
\tikzstyle{doubleArrow} = [thick, decoration=
	{markings,
	mark=at position 0 with {\arrow[semithick]{open triangle 60 reversed}},
	mark=at position 1 with {\arrow[semithick]{open triangle 60}}},
   double distance=1.4pt, shorten >= 5.5pt,
   preaction = {decorate},
   postaction = {decorate}]
\tikzstyle{innerWhite} = [semithick, white,line width=1.4pt, shorten >= 4.5pt]
\pgfplotsset{
    colormap={blackwhite}{[5pt]
        rgb(0pt)=(1, 0, 0);
        rgb(1000pt)=(0, 0, 1)
    },
}
\tikzset{dot/.style={circle,fill=#1,inner sep=0,minimum size=4pt}}

\usepgfplotslibrary{fillbetween}
\pgfplotsset{compat=1.10}
\tikzset{>=latex}

\tikzset{
  markergrau/.style={
    rectangle,
    fill = himmel,
    rounded corners,
    minimum height=2.5em,
    minimum width=6em,
    inner sep=5pt,
    text centered,
  }
}

\tikzset{
  markerspecial/.style={
    rectangle,
    fill = grey,
    rounded corners,
    minimum height=2.5em,
    minimum width=6em,
    inner sep=5pt,
    text centered,
  }
}

\tikzset{
  markerellipse/.style={
    ellipse,
    fill = grey,
    minimum height=2.5em,
    minimum width=6em,
    inner sep=0pt,
    text centered,
  }
}

\tikzset{
  markernone/.style={
    draw = none,
    fill = none
  }
}

\tikzset{
  randbox/.style={
    rectangle,
    fill = darkgrey,
    minimum height=3cm,
    minimum width=2em,
    inner sep=5pt,
    text centered,
  },
}

\tikzset{ defines/.style={ -, thick } }
\tikzset{ triggers/.style={ ->, thick } }
\tikzset{ triggersOut/.style={ -|, thick } }
\tikzset{ triggersIn/.style={ |->, thick } }
\tikzset{ impacts/.style={ ->, dotted, thick } }
\tikzset{ blocks/.style={ -|, double, thick }, }
\tikzset{ arrlabel/.style={right=0.15cm} }
\tikzset{ arrlabell/.style={left=0.15cm} }
\tikzset{ arrlabelu/.style={above=0.15cm} }
\tikzset{ arrlabelb/.style={below=0.15cm} }
\newcommand{\mymarkerexp}[3]{
  \node[#3] (#2) {
    \textbf{ \begin{tabular}{c} #1 \end{tabular} }
  }; }

\newcommand{\mymarker}[2]{ \mymarkerexp{#1}{#1}{#2} }

\newcommand{\mytrigger}[3]{ \path (#1) edge[triggers] node[arrlabel]{#2} (#3); }

\newcommand{\mytriggerlIn}[3]{ \path (#1) edge[triggersIn] node[arrlabell]{#2} (#3); }
\newcommand{\mytriggerl}[3]{ \path (#1) edge[triggers] node[arrlabell]{#2} (#3); }

\newcommand{\mytriggerlOut}[3]{ \path (#1) edge[triggersOut] node[arrlabel]{#2} (#3); }

\newcommand{\mytriggeru}[3]{ \path (#1) edge[triggers] node[arrlabelu]{#2} (#3); }

\newcommand{\figuresAppendix}[7]{
\begin{figure*}
\begin{center}
\begin{subfigure}[c]{0.71\textwidth}
\begin{center}
\includegraphics[width=1.\textwidth]{#1} 
\end{center}
\end{subfigure}

\bigskip
\bigskip
\begin{subfigure}[c]{0.71\textwidth}
\begin{center}
\includegraphics[width=1.\textwidth]{#2}
\end{center}
\end{subfigure}

\caption{Optimization results for Patient~#4 (top, #6) and #5 (bottom, #7).} 
\label{#3}
\end{center}
\end{figure*}
}

\usepackage{lastpage,fancyhdr,graphicx}
\usepackage{epstopdf}

\usepackage{atenddvi}
\usepackage[user]{zref}

\newcounter{pageaux}
\def\currentauxref{PAGEAUX1}

\makeatletter
\newcommand{\resetpageaux}{%
  \clearpage
  \edef\@currentlabel{\thepageaux}\label{\currentauxref}%
  \xdef\currentauxref{PAGEAUX\thepage}%
  \setcounter{pageaux}{0}}
\AtEndDvi{\edef\@currentlabel{\thepageaux}\label{\currentauxref}}
\makeatletter

\ifCLASSINFOpdf
\else
\fi
\begin{document}
\title{Model-based optimal AML consolidation treatment}

\author{Felix Jost, Enrico Schalk, Daniela Weber, Hartmut D\"ohner, Thomas Fischer \& Sebastian Sager% <-this % stops a space

\thanks{F. Jost and S. Sager are with the Otto-von-Guericke University, Department of Mathematics, Magdeburg, Germany.}
\thanks{E. Schalk and T. Fischer are with the University Hospital
Magdeburg, Department of Hematology and Oncology, Magdeburg, Germany.}
\thanks{D. Weber and H. D\"ohner are with the University Hospital Ulm, Department of Internal Medicine III, Ulm, Germany}
\thanks{E. Schalk, T. Fischer and S. Sager are with the Health Campus Immunology, Infectiology and Inflammation (GC-$\text{I}^3$), Medical Faculty, Otto-von-Guericke University, Magdeburg, Germany}
\thanks{Manuscript received November 27, 2019;revised January 8, 2020; accepted March 17, 2020. This project has received funding from the European Research Council (ERC, grant agreement No 647573) and from the European Regional Development Fund (grants SynMODEST and SynIsItFlutter) under the European Union's Horizon 2020 research and innovation program. (Corresponding author: Sebastian Sager)}
\thanks{Copyright (c) 2017 IEEE. Personal use of this material is permitted. However, permission to use this material for any other purposes must be obtained from the IEEE by sending an email to pubs-permissions@ieee.org.}}


\markboth{}%
{Shell \MakeLowercase{\textit{et al.}}: Bare Demo of IEEEtran.cls for IEEE Journals}



\maketitle

\begin{abstract}
Objective:
Neutropenia is an adverse event commonly arising during intensive chemotherapy of acute myeloid leukemia (AML). It is often associated with infectious complications.
% Ziel
Mathematical modeling, simulation, and optimization of the treatment process would be a valuable tool to support clinical decision making, potentially resulting in less severe side effects and deeper remissions. 
However, until now, there has been no validated mathematical model available to simulate the effect of chemotherapy treatment on white blood cell (WBC) counts and leukemic cells simultaneously.
Methods:
% model erklaere
We developed a population pharmacokinetic/pharmacodynamic (PK/PD) model combining a myelosuppression model considering endogenous granulocyte-colony stimulating factor (G-CSF), a PK model for cytarabine (Ara-C), a subcutaneous absorption model for exogenous G-CSF, and a two-compartment model for leukemic blasts. 
This model was fitted to data of 44 AML patients during consolidation therapy with a novel Ara-C plus G-CSF schedule from a phase II controlled clinical trial.
Additionally, we were able to optimize treatment schedules with respect to disease progression, WBC nadirs, and the amount of Ara-C and G-CSF. 
% daten erklären
Results:
% numerische ergebnisse simulationen erklaeren
The developed PK/PD model provided good prediction accuracies and an interpretation of the interaction between WBCs, G-CSF, and blasts. 
For 14 patients (those with available bone marrow blast counts), we achieved a median 4.2-fold higher WBC count at nadir, which is the most critical time during consolidation therapy. 
The simulation results showed that relative bone marrow blast counts remained below the clinically important %relapse-free 
threshold of 5\%, with a median of 60\% reduction in Ara-C.
Conclusion:
These in silico findings demonstrate the benefits of optimized treatment schedules for AML patients.
Significance:
Until 2017, no new drug had been approved for the treatment of AML, fostering the optimal use of currently available drugs.
\end{abstract}

\begin{IEEEkeywords}
Myelosuppression, Population PK/PD modeling, cytarabine, lenograstim, treatment schedule optimization.
\end{IEEEkeywords}

% \fi
\IEEEpeerreviewmaketitle

\section{Introduction}
\IEEEPARstart{C}{hemotherapy} treatment of acute myeloid leukemia (AML) is usually divided into an induction phase and a consolidation phase.
The goal of the induction phase is the eradication of blasts. By blasts, we refer to a combination of aberrant/leukemic and physiological blasts that are cytologically $\ge$ 20\% in the bone marrow (BM) at the time of AML diagnosis \cite{Doehner2017}.
The standard treatment consists of intensive chemotherapy with three days of anthracycline (idarubicin or daunorubicin) and seven days of cytarabine (Ara-C).
To monitor the relative numbers of blasts in each cycle, BM aspirations are collected and analyzed.
After the induction phase, for a hematological complete remission the relative number of blasts should be below 5$\%$ in the BM (assessed by cytology) and not measurable in the circulating blood \cite{De2005}.
In this study, we are interested in the subsequent consolidation phase. 
The phase consists of up to four consolidation cycles (CCs) of intermediate- or high-dose Ara-C or allogeneic hematopoietic stem cell transplantation \cite{Doehner2017}.
While the goal to reduce the blasts as much as possible to prevent a relapse is identical between the two phases, the conflicting objective to avoid complicating infections plays an important role in the consolidation phase. 
Neutropenia, defined by the common terminology criteria for adverse events \cite{Us2018}, is characterized by decreased counts of neutrophil granulocytes in the peripheral blood. It is a serious and common adverse event arising during the treatment with cytotoxic chemotherapy of AML.
This form of white blood cell (WBC) suppression in the BM (myelosuppression) is responsible for a higher risk of infections and consequently for delayed, dose-reduced or stopped treatments, longer hospitalization periods, and mortality as the worst case.
In this work, we focus on grade 4 leukopenia (WBC count $<$ 1 G/L) \cite{Us2018} in the consolidation phase.
In clinical practice, grade 4 leukopenia is equivalent to grade 4 neutropenia (absolute neutrophil count $<$ 0.5 G/L) \cite{Us2018} and was chosen because the available measured WBC counts were not further specified into granulocytes (neutrophils, eosinophils, and basophils), monocytes, or lymphocytes (T cells and B cells).
In each cycle of cytotoxic chemotherapy, the WBCs decrease from their normal range to a critical value close to 0 G/L.
The interval in which the WBCs are below a certain grade is defined as leukopenia, and the time from the start of treatment until WBC recovery above this threshold is called WBC recovery time. 
One standard AML consolidation treatment consists of Ara-C 3 $g/m^2$ (D) intravenous (body surface area [BSA]-adjusted) lasting 3 hours every 12 hours on days 1, 3 and 5 (D135) for patients aged 60 years and younger, which was investigated by Mayer \textit{et al.} in 1994 \cite{Mayer1994}. 
Older patients ($>$ 60 years) receive an intermediate dosage of 1 $g/m^2$ Ara-C (d) infusions in the same intervals (d135).
In recent years, studies have proposed a dense treatment schedule at which either high-dose or intermediate-dose Ara-C is administered on days 1, 2, and 3 (D123 or d123) to reduce the WBC recovery time and increase survival \cite{Jaramillo2017, Schlenk2019}. 

In addition to new treatment schedules, the administration of granulocyte-colony stimulating factor (G-CSF) can reduce the depth and duration of leukopenia \cite{Paul2014}.
Hematopoietic growth factors such as G-CSF regulate blood cell production, including survival, proliferation, and differentiation of hematopoietic stem cells and stimulation of mature cell functions by activating signal transduction pathways \cite{Hoffbrand2005}.
The impact of G-CSF was enhanced by the clinical development of a recombinant human G-CSF, called filgrastim, in 1986 for the prevention of leukopenia and hematopoietic stem cell mobilization before autologous or allogeneic hematopoietic stem cell transplantation \cite{Welte1996}.
The European Society for Medical Oncology suggests daily filgrastim administration after the last day of chemotherapy until a sufficient/stable postnadir absolute neutrophil count recovery, respectively, for approximately 10 days \cite{Klastersky2016}. % https://www.accessdata.fda.gov/drugsatfda_docs/label/2012/103353s5147lbl.pdf
On top of chemotherapy, the additional burden of daily filgrastim administration was reduced by the invention of pegfilgrastim, a pegylated form of filgrastim.
The inclusion of filgrastim into a polyethyleneglycol polymer prolongs the half-life from 3.5 hours to 46-62 hours such that the permanence of pegfilgrastim in blood circulation is up to 16 days after a single administration \cite{Bolis2013}, replacing the frequent filgrastim administrations.
Also lenograstim was developed, which is a physicochemically, immunologically and biologically identical glycosylated recombinant G-CSF to human G-CSF \cite{Bolis2013}.
Subcutaneous (s.c.) lenograstim has a bioavailability of 30\% with a dose-dependent decrease from 62\% to 24\% over the dose range of 0.5-10 $\mu g/kg/day$ and an elimination half-life of 3-4 hours at steady state (repeated dosing) \cite{Fernandez2007}.
%%%%%%%  Purpose of model development
%In this work we use pharmacokinetic(PK)/pharmacodynamic(PD) modeling approaches to describe clinical data 
The main contribution of this work is the formulation of a pharmacokinetic (PK)/pharmacodynamic (PD) model that describes the dynamics of WBCs and leukemic cells under consideration of Ara-C and external G-CSF administration and a comprehensive \textit{in silico} analysis of personalized models using optimal control technology. 
In particular, we used nonlinear mixed-effects PK/PD modeling approaches together with clinical data to explore the impact of different treatment schedules and the administration of lenograstim on WBCs and leukemic blasts. % and we propose individually optimized treatment schedules.
In recent years, PK/PD models for endogenous G-CSF and blood cells such as neutrophils or leukocytes \cite{Quartino2014} and several PK models for exogenous G-CSF (filgrastim \cite{Wang2001,Foley2009,Wiczling2009,Krzyzanski2010}, pegfilgrastim \cite{Roskos2006,Foley2009,Yang2011} and lenograstim \cite{Hayashi1999,Akizuki2000,Hayashi2001,Ria2010,Kagan2014}) have been presented. 
However, few PK/PD models combining endogenous and exogenous G-CSF and leukocytes \cite{Scholz2012,Schirm2018} or neutrophils \cite{Shochat2008,Brekkan2018,Melhem2018} for the prediction of myelosuppression have been published, %Interaction between neutrophils and leukemic cells \cite{Rubinov1976,Afenya1996, Stiehl2012, Pefani2014, }
particularly not for Ara-C-derived myelosuppression during consolidation therapy of AML patients.
Therefore, we developed a population PK/PD model modifying, and extending previously published models to analyze the inverse correlation between G-CSF and leukocytes during different Ara-C and lenograstim schedules.
The consideration of leukemic blasts and their interaction with leukocytes, comparable to previous works \cite{Rubinov1976,Afenya1996, Stiehl2012}, completed the model.
The resulting mathematical model is to our knowledge the first one which can describe the dynamics of WBC, leukemic cells, and G-CSF at the same time for the CC of AML patients.
%We personalized the model via regression to clinical data and calculated mathematically optimal solutions, extending the analysis hence beyond previous results for isolated submodels \cite{Djulbegovic1985,Matveev2001,Pefani2014} to a comprehensive view on Ara-C plus lenograstim consolidation treatment.
% formulated a multiobjective optimization problem considering terms for disease progression, state of health, and therapy costs.
%
%In summary, we developed models and algorithms for a computational framework to individually simulate, analyze, and optimize the consolidation treatment schedules of Ara-C and lenograstim.
We investigated sensitivities of the numerical results with respect to the choice of submodels, but also with respect to the choice of the objective function of the optimal control problem, which is well known to have a strong impact on optimal solutions \cite{Engelhart2011}.

 

\section{Patients and methods}

\subsection{Patients and clinical data}

Data from the AMLSG 12-09 randomized controlled clinical phase II trial \cite{Schlenk2019} were provided by the Department of Internal Medicine III, University Hospital Ulm, Ulm, Germany and used for model development, fitting, validation, and calibration.
Patients were randomized to one out of four different induction therapies following either allogeneic hematopoietic cell transplantation or three cycles of high-dose Ara-C. 
The dataset (denoted by Ulm in Figure~\ref{fig:VPC_GoF_Cycles}(a)) included WBC count measurements (6-16 per cycle) from 86 Ara-C CCs, partitioned into one, two, and three consecutive CCs from 20, 6, and 18  AML patients (median 65 years, 19 [43\%] male, median 1.8 BSA [$m^2$]), respectively, from 2010 and 2012, which were treated with D123 (31 out of 86 CCs) or d123 (55 out of 86 CCs) schedules of Ara-C. 
Additionally, in most cycles before Ara-C treatment (76 measurements), the relative number of blasts in the BM and the category of BM cellularity (punctio sicca, hypo-, normo- or hypercellular) were determined by cytology via BM aspiration.
13 BM measurements were below the limit of quantification and consequently excluded from the analysis.
The treatment schedule included $263 \ \mu g$ of s.c. lenograstim administrations (one patient received $324 \ \mu g$ in one CC) starting nine days after the start of  Ara-C treatment until hematological recovery, i.e., neutrophil count $>$ 0.5  $G/L$, was achieved.
Nine of the patients (1 only in the first cycle and 7 CCs each for d123 and D123) did not receive lenograstim.
For the analysis in the subsequent section called \textit{Modeling exogenous G-CSF}, the patientwise cycles of the current dataset were treated independently (although several cycles belong to the same patient) and combined with the publicly available dataset (denoted by MD in Figure~\ref{fig:VPC_GoF_Cycles}(a)) from the supporting information of \cite{Jost2019}.
This dataset was retrospectively collected from records of clinical routine and provided by the Department of Hematology and Oncology, Magdeburg University Hospital, Magdeburg, Germany.
The dataset consists of one, two, and three consecutive CCs from nine, nine, and five patients (median 62 years, 14 [61\%] male, median 1.8 BSA [$m^2$]), respectively, who received different treatment schedules of D135, d135, D123 and D12.
In the section titled \textit{Modeling leukemic blasts} the current dataset was used for model fitting.
In the following section \textit{Model predictions and optimal treatment schedules}, a subset of 24 patients, for whom at least two CCs were available, were used to perform model predictions.
For the computation of optimized treatment schedules, this subset was further reduced to 14 patients for whom relative blast counts were available in the last CC.
The different subsets used in each section are visualized as a diagram in~\ref{fig:FlowChartStudies} in the supplemental material (SM).

\subsection*{PK/PD model}

%\begin{enumerate}
%\item parameterized secondary Ara-C PD effect (no cancer cells)
%\item full model with cancer cells fitted to Ulm data
%\item predicting last consolidation cycle
%\item optimizing schedule of last cycle
%\end{enumerate}

\begin{figure*}[!t]
\begin{center}
\resizebox{1.\textwidth}{!}{
\begin{tikzpicture}[->,>=stealth']
  \mymarkerexp{Proliferating cells \\ $\mxpr$}{xpr}{markergrau}
  \mymarkerexp{Transit cells \\ $\mxtr$}{xtr}{markergrau, below=1cm of xpr}
  \mymarkerexp{Mature WBC \\ $\mxma$}{xma}{markergrau, below=1cm of xtr}
  \mymarkerexp{}{death}{markernone, below=0.6cm of xma}
  \mymarkerexp{}{feedback}{markernone, left=0.8cm of xpr, yshift=-1.3cm}
  \mymarkerexp{}{feedbackbeta1}{markernone, left=-2.4cm of xpr, yshift=-1.2cm}
  \mymarkerexp{}{feedbackbeta2}{markernone, left=-1.8cm of xtr, yshift=-1.2cm}
  \mymarkerexp{}{feedback2}{markernone,  above left=0.8cm of xpr, yshift=-1.3cm}

  % endogenous GCSF
  \mymarkerexp{endogenous GCSF \\ $\mxG$}{xG}{markergrau,  left=3cm of xtr}
  \mymarkerexp{}{deathGCSF}{markernone, below=0.6cm of xG}
  \mymarkerexp{}{deathGCSF2}{markernone, below right=0.7cm of xG}
  \mymarkerexp{}{prodGCSF}{markernone, above=0.6cm of xG}
  \mymarkerexp{}{Dummyfeedback}{markernone, above right=3.cm and -1.cm of xG}
  \mymarkerexp{}{death3}{markernone, above=0.6cm of xG}
  \mytriggerl{$(death3)-(1,0)$}{$\mkin$}{$(xG.north)-(1,0)$}
  %\mytriggerl{xG}{$\mkin$}{prodGCSF}
  \mytrigger{xG}{$\mkoutLeukemic$}{deathGCSF}

  \mytrigger{xpr}{$\mktr$}{xtr}
  \mytrigger{xtr}{$\mktr$}{xma}
  \mytrigger{xma}{$\mkma$}{death}
  \path[impacts] (xG) edge[bend left=31] node[below] {$\feedbackBeta$}  (feedbackbeta1);
  \path[impacts] (xG) edge[bend left=10] node[above] {$\feedbackBeta$} (feedbackbeta2) ;
  \path[impacts] (xG) edge[bend left=30] node[above, rotate=45] {$\feedbackGamma$} (feedback2);

  %\path[triggers] (xpr) edge[loop left above] node[pos=0.15,below,sloped] {$\mfunFcomp$} ();
  \draw [triggers] (xpr.west) arc (-45:-355:8mm) node[pos=0.75,above] {$\mkpr = \mktr$};

%  \path[impacts] (xma) edge[bend right=31] (feedbackbeta1);
%  \path[impacts] (xma) edge[bend right=31] (feedbackbeta2);

% ---
  \mymarkerexp{Ara-C (peripheral) \\ $x_\textbf{2}$}{x2}{markergrau, above=2cm of xpr}
  \mymarkerexp{Ara-C (central) \\ $x_\textbf{1}$}{x1}{markergrau, left=1cm of x2}
  \mymarker{E}{markerellipse, below=0.7cm of x1}
  \mymarkerexp{}{death2}{markernone, above=0.6cm of x1}

  \path[triggers] (x1.east) edge[bend left=31] node[pos=0.5,above,sloped] {$\mkab$} (x2.west);
  \path[triggers] (x2.west) edge[bend left=31] node[pos=0.5,below,sloped] {$\mkba$} (x1.east);
  \mytrigger{$(x1.north)+(1,0)$}{$\mkaz$}{$(death2)+(1,0)$}
  \mytriggerl{$(death2)-(1,0)$}{Ara-C ($u_c$)}{$(x1.north)-(1,0)$}
  \mytriggerl{x1}{\mslope}{E}
  \path[triggersOut] (E) edge[bend right=31] ($(feedback2)+(-.6,.3)$) ; % (feedback2);
  %\path[triggersOut] (E) edge[bend right=61] (Dummyfeedback);

  \path[impacts] (xma) edge[bend left=30] node[right] {} (deathGCSF2) ;

  % cancer cells
  \mymarkerexp{Leukemic blasts (bone marrow) \\ $x_{\textbf{l}_\textbf{1}}$}{l1}{markergrau, left=5cm of x1}
  \mymarkerexp{}{dummyBlasts}{markernone, below=1cm of l1}
  \mymarkerexp{Leukemic blasts (blood) \\ $x_{\textbf{l}_\textbf{2}}$}{l2}{markergrau, below=1cm of dummyBlasts}
  \mymarkerexp{}{deathLeukemic}{markernone, below=0.6cm of l2}
  \path[impacts] (l2) edge[bend right=40] node[right] {} (deathGCSF2) ;
  %\draw [triggers] (l1.east) arc (+180:+505:8mm) node[pos=0.75,above] {$2a_1p_1$};
  \mymarkerexp{}{feedbackGcsfCancer}{markernone, below=1.cm of l1, xshift=+2.5cm}
  \path[impacts] (l2.east) edge[bend right=80] node[right] {$k_{lc} = \feedbackGcsfCancer$} (feedbackGcsfCancer) ;
  \path[impacts] (xma) edge[bend right=20] node[right] {} (feedbackGcsfCancer) ;
  %\mytrigger{l1}{$\mktr$}{l2}
  \mymarkerexp{}{in3}{markernone, above=0.8cm of l2}
  \mytriggerlOut{l1}{$p_1$}{dummyBlasts} 
  \mytriggerlIn{$(dummyBlasts)+(0,0.25)$}{$2(1-a_1k_{lc})p_1$}{$(l2.north)-(0,0)$}
  %\mytriggerlOut{$(in3)+(1,0)$}{$p_1$}{$(l2.north)+(1,0)$}
  \mytriggerl{l2}{$d_2$}{deathLeukemic}
  \path[triggersOut] (E) edge[bend right=0] ($(in3)+(1.,.9)$);
  \path[triggers] ($(dummyBlasts)+(0.,0.25)$) edge[bend left=-80] node[pos=0.5,below,sloped] {$2p_1a_1k_{lc}$} (l1.east);
  %\path[impacts] (E) edge[bend left=11] ($(feedbackGcsfCancer)+ (0.4,-0.1)$);

  % exogenous GCSF
  %\mymarkerexp{exo. GCSF (central) \\ $\mxSCG$}{xSCG}{markergrau,  below=3cm of xG};
  \mymarkerexp{Transit Comp. \\ $x_{\textbf{exo}_\textbf{2}}$}{xG3}{markergrau,  markergrau,  below=1.5cm of l2};
  \mymarkerexp{Transit Comp. \\ $x_{\textbf{exo}_\textbf{1}}$}{xG2}{markergrau,  markergrau,  below=1.cm of xG3};
  \mymarkerexp{Lenograstim \\ Depot \ $x_\textbf{D}$ }{xPerG}{markergrau,  right=1.5cm of xG2};

  %\mymarkerexp{exo. GCSF (peripheral) \\ $\mxCPG$ }                      {xMMG}{markergrau,  left=1cm of xSCG};
  %\path[triggers] (xSCG.west) edge[bend left=31] node[pos=0.5,below,sloped] {$\mkabGCSF$} (xMMG.east);
  %\path[triggers] (xMMG.east) edge[bend left=31] node[pos=0.5,above,sloped] {$\mkbaGCSF$} (xSCG.west);
  %\path[triggers] (xSCG.north) edge[bend left=31] node[pos=0.5,above,sloped] {$\mka$} (xG.west);
  \mymarkerexp{}{prodExGCSF}{markernone, right=0.6cm of xPerG};
  \mymarkerexp{}{absorpExGCSF}{markernone, right=0.6cm of xPerG};
  \mymarkerexp{}{absorpExGCSF2}{markernone, right=1.5cm of xPerG};
  \mytrigger{prodExGCSF}{Lenograstim ($u_l$)}{xPerG};
  \mytriggeru{xPerG.west}{$k_{a1}$}{xG2.east};
  \mytrigger{xG2.north}{$k_{a2}$}{xG3.south};

  \mytriggeru{xG3.east}{$k_{a2}$}{xG};

\end{tikzpicture}}
\end{center}
\caption{Visualization of the final pharmacokinetic/pharmacodynamic model. The hematopoiesis of white blood cells (WBCs) is described by two compartments representing the proliferation and differentiation within the bone marrow. The third compartment describes the circulating matured WBCs. 
The sequential hierarchy (similar to WBC) of leukemic blasts is described by a two-compartment model.
Both linages interact by the competition of endogenous G-CSF.
Ara-C affects proliferation of leukemic blasts and WBCs.
Lenograstim administration was modeled by a single pathway absorption model with two transit compartments \cite{Kagan2014}.}
\label{fig:SchemeCancer}
\end{figure*}
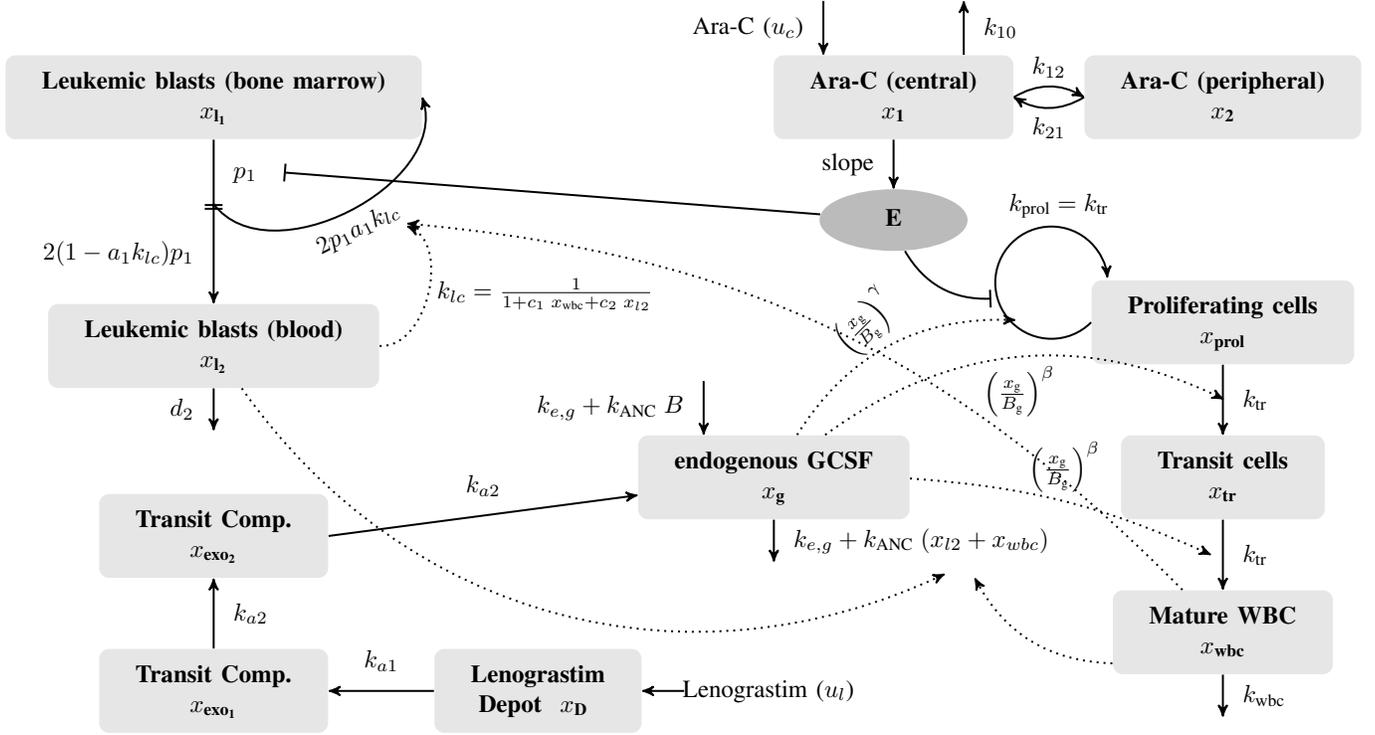

The aim was to develop a population PK/PD model describing WBC counts and blasts of AML patients treated with Ara-C and lenograstim during consolidation therapy.
The model development was guided by previously published models and available WBC counts and blast measurements.
The PK/PD model by Quartino \textit{et al.} \cite{Quartino2014} describing the proliferation and differentiation of stem cells to mature neutrophils, and its regulation by endogenous G-CSF was used as a starting point.
%In a first step, the myelosuppression model was expanded with a PK model of Ara-C \cite{Jost2019} and a single pathway model with transit compartments decribing the subcutaneous administration of lenograstim \cite{Kagan2014}.
%The model was fitted to a variety of different consolidation cycles.
%In a second step, a two-compartment model describing the dynamics of leukemic blasts was incorporated.
The developed PK/PD model is shown in Figure \ref{fig:SchemeCancer}.

%%%% myelosuppression model 
The two-compartment PK model ($x_1,x_2$) for Ara-C was taken from \cite{Jost2019} as our clinical data did not contain Ara-C measurements.
The PK model describes the PKs and biphasic elimination of Ara-C after high-dose infusions \cite{Kern1997}.
% and linked to the model from \cite{Quartino2014}.
The hematopoiesis of WBCs is modeled by a chain of three compartments representing the proliferating stem cells $\mxpr$ and differentiating cells $\mxtr$ in the BM released to the blood stream after maturation to WBCs.
Matured cells $\mxma$ die by apoptosis with a death rate constant $\mkma$.
Ara-C is incorporated into the DNA leading to cell death, such that a log-linear PD term as a first order kinetics negatively influences the proliferation of stem cells, equivalent to \cite{Jost2019}.   
We assumed that the plasma concentration is an adequate surrogate for the PD effect of Ara-C in the BM, as no PK model for high-dose Ara-C is available which considering the BM as an additional compartment.
We take up this assumption in the discussion.
%$\mxpr$ and $\mxma$ represent the proliferating cells in the BM and circulating mature WBCs in the blood stream.
Two modifications of the myelosuppression model were implemented for our purposes.
Instead of three, we used one transit compartment $\mxtr$, still guaranteeing a reliable interpretation of the mean maturation time and no loss of model accuracy. 
A detailed discussion is given in \cite{Jost2019}.
%Furthermore, a second PD term was added to the proliferation speed parameter $\mgamma$, as discussed in the next section.
%%%% GCSF
Furthermore, the s.c. administration of lenograstim was modeled by a chain of three compartments describing the effect of enhanced proliferation and maturation.
%The first compartment is a depot compartment with the constant F representing the bioavailability of s.c. administration of lenograstim which was determined to be 30\% \cite{Fernandez2007}.
As lenograstim has an equivalent chemical structure than endogenous G-CSF and they bind to the same receptors \cite{Bolis2013}, lenograstim is released to compartment $x_g$ via the first-order absorption rate constants $k_{a1}$ and $k_{a2}$ positively affecting the production of WBCs.
As no endogenous G-CSF measurements were available, several parameters were fixed to values from publications (see~\ref{tab:ModelSpecifications}).
The leukemic blasts were included as a separate cell line.
The sequential hierarchy \cite{Bonnet1997,Hope2004} of leukemic blasts (similar to WBC) is described by the two compartments $x_{l1}$ and $x_{l2}$ which represent the leukemic blasts in the bone marrow, respectively circulating blood and was published in \cite{Stiehl2018}.
The leukemic blasts in the bone marrow grow and proliferate with the first order rate $p_1$.
During cell division a leukemic blast divides into two daughter cells, so that the outflux from mitosis is $2p_1x_{\textbf{l}_\textbf{1}}$.
The outflux is then separated into the process of self-renewal by the rate $2p_1a_1k_l$ with the fraction constant $a_1$ determining the fraction of daughter cells staying at the current differentiation stage and cell movement by the rate $2p_1(1-a_1)$ to the consecutive compartment. 
Leukemic cells are dying by the first order rate $d_2$.
%We used the two compartment model from Stiehl to describe the sequential hierarchy (similar to WBC) leukemic cells are organized \cite{Bonnet1997,Hope2004}. % ref from Stiehl2018 
In contrast to the myelosuppression model, which does not distinguish between self-renewal and differentiation into the next compartment, the model of the leukemic blasts takes this separation into account.
As we concentrated on the cytokine-dependent version of leukemic blasts, we used the term $k_{lc}$ from \cite{Stiehl2012,Stiehl2018} in which the interaction between leukemic blasts and WBC counts is modeled through the competition of endogenous G-CSF between the circulating cells of both lineages.
The term was derived from a quasi-steady-state assumption of the G-CSF dynamics (see \cite{Stiehl2012} for a detailed discussion).
The WBC-regulated elimination of G-CSF $k_{out}$ was extended with circulating leukemic blasts $x_{l2}$ because both linages make use of G-CSF.
The interplay between the G-CSF induced proliferation of leukemic and healthy cells drives the system into a purely leukemic or healthy steady state \cite{Stiehl2012}.
A numerical steady state analysis was performed to determine the system behavior until one and a half years after the start of the first CC. 
For each patient, the validated model drives into a purely leukemic steady state ($\mxpr = \mxtr = \mxma = 0$ and $x_{l1} , x_{l2} > 0$) after five months on average.
Values for $a_1$ and $d_2$ were taken from \cite{Stiehl2018} and $p_1$ was chosen as a half of the WBC proliferation similar to \cite{Stiehl2018} characterizing slow growing leukemic cells \cite{Nobile2019} resulting in a duration of remission in the range of 4.1 to 8.1 months reported by \cite{Cassileth1988}.
To formulate the mathematical model as a system of ordinary differential equations in a compact form, we use the following definitions, i.e., for the PD effect $E$ on $\mktr$, the zero-order production rate $k_{in}$ and first-order elimination rate $k_{out}$ of G-CSF, and the G-CSF quasi-steady-state term $k_{lc}$ for the leukemic blasts, we define
\begin{linenomath*}
\begin{align*}
E &=\mslope \; ln(\frac{x_1}{V_c \; \mMM} + 1) \\
%S &= (1 + \mslopeGamma \; ln(\frac{x_1}{V_c \; \mMM} + 1)) \\
k_{in} &= ( k_{e,g} + k_{\text{ANC}} \; B ) \; B_g +  \ k_{a2} \ x_{\text{exo}_1} \\
k_{out} &= ( k_{e,g} + k_{\text{ANC}} \; ( \mxma + x_{l2}  )) \\
k_{lc} &= \frac{1}{1+ c_1 \; \mxma + c_2 \; x_{l2}}.
\end{align*}
\end{linenomath*}
All constants, control functions defining the administration of Ara-C and lenograstim, parameters, and initial conditions are specified in~\ref{tab:ModelSpecifications} in the SM.
The time derivatives of all states are given by%
\begin{linenomath*}
\begin{subequations} \label{eq:ODEmodel}
\begin{align} 
\dot x_{1} &= - (\mkaz+\mkab) \; x_1 + \mkba \; x_2 + \muExpr \\
\dot x_2   &= \hspace{9.7ex} \mkab \; x_1 - \mkba \; x_2 \\
\dot x_{prol} &= - \feedbackBeta \mktr \; \mxpr + \feedbackGamma \mktr (1-E) \; \mxpr \label{eq_bloodc} \\
\dot x_{tr} &= \phantom{-} \feedbackBeta \mktr \; \mxpr - \feedbackBeta \mktr \; \mxtr \label{eq:xtr} \\
\dot x_{\text{wbc}} &=  \phantom{-} \feedbackBeta \mktr \; \mxtr - \mkma \ \mxma \label{eq:xma} \\
\dot x_g &= k_{in} - k_{out} \; x_g  \\
\dot x_D &= -k_{a1} \; x_D + \frac{u_{l} \; 1000}{V_g \; \text{dur}_{l}} \\
\dot x_{\text{exo}_1} &= k_{a1} \; x_D - k_{a2} \; x_{\text{exo}_1} \\
\dot x_{\text{exo}_2} &= k_{a2} \; x_{\text{exo}_1} - k_{a2} \; x_{\text{exo}_2}  \\ 
\dot x_{l1} &= (2a_1 k_{lc} - 1) p_1 \; x_{l1} - p_1 E \;  x_{l1}  \\ 
\dot x_{l2} &=  2(1-a_1 k_{lc})p_1 \; x_{l1}  -d_2 \; x_{l2}.
\end{align}
\end{subequations}
\end{linenomath*}

\begin{table*}[!t]
\caption{Model constants, patient-specific constants, model parameters, and initial values with their units and descriptions.
%The constants of the PK model for Ara-C were determined from published data \cite{Kern1997, Jost2019} and applied to all patients.
} 
\label{tab_Constants}
\centering \footnotesize
\bgroup
\def\arraystretch{.9}
\setlength\tabcolsep{3.5pt}
%\resizebox{1.\textwidth}{!}{
\begin{tabular}{|lcrl| }
\toprule
\rowGrey \textbf{Constant} & \textbf{Unit} & \textbf{Value} & \textbf{Description}\\
%  Constant & Unit & Value & Description\\
%\midrule
  \multicolumn{4}{|l|}{ \textbf{PK model of Ara-C}} \\ 
$\mkaz$          &  $1/day$  & $98.2920$  & Elimination rate of Ara-C\\
$\mkab$          &  $1/day$  & $2.6616$   & Distribution rate of Ara-C  \\
$\mkba$          &  $1/day$  & $12.8784$  & Distribution rate of Ara-C  \\
Volume $\mV$                &  $L$  & $37.33$ & Volume of central compartment  \\ % $37.33$ \\
$\mMM$       &  $g/mol$  & $243.217$  & Molecular mass of Ara-C \\
$dur_c$       &  $day$  & $1/8$  & Infusion time \\
  \multicolumn{4}{|l|}{\textbf{PD model of WBCs and leukemic blasts}} \\ 
$\mkma$            &  $1/day$    & $2.3765$ & Death rate of circulating WBCs \\
$\beta$            &     -       &  0.234   &   Feedback of G-CSF on transit time   \\
$\mslopeGamma$     &     -       &  0.470  & PD effect of Ara-C on $\mgamma$      \\
$B_g$              &   $ng/L$    &     24.4   & Endogenous G-CSF steady state   \\
$k_{e,g}$            &  $1/day$    &    0.592$\times$24  & Non-specific elimination rate constant \\
$k_{ANC}$          &  $1/day$    &    5.64$\times$24 & Neutrophil-dependent elimination rate \\
$a_1$              &     -       & 0.875  &  Probability of self-renawal  \\
$p_1$              &     $1/day$ & 0.1  &  Leukemic cell proliferation rate  \\
$d_2$              &     $1/day$ & 2.3  &  Leukemic cell death rate \\
$c_1$              &   $L/10^9$  &  0.01   &   G-CSF quasi steady-state feedback scaling factor \\ % constant     \\
$c_2$              &   $L/10^9$  &  0.01  &   G-CSF quasi steady-state feedback scaling factor \\ % constant      \\
$k_{ANC}$              &  $1/day$    & 5.64$\times$24  & ANC dependent elimination rate  \\
                   &                 &           & of endogenous G-CSF \\
  \multicolumn{4}{|l|}{\textbf{PK model of lenograstim}} \\ 
F                  &  -    & 0.3  &  Bioavailability of s.c. administration from \cite{Fernandez2007}  \\
$V_g$                       &  $L$         &  14.5 & Volume of distribution from \cite{Hayashi1999}  \\ % \midrule
$dur_l$       &  $day$  & $0.0007$  & Infusion time \\
$k_{e,exo}$              &  $1/day$    & 0.220$\times$24  & Elimination rate of exogenous G-CSF \\
%\midrule \rowGrey Patient-Specific &  Unit  & Range  & \\ \midrule
$\mBSA$            &  $m^2$  & $[1.61,2.07]$ & Body surface area \\
$u_{c}(t)$             &  $g/m^2$ & $[1,3]$  & Ara-C $\mdosage$ \\
$u_{l}(t) $     &  $\mu g$ & $[263,324]$  & Lenograstim $\mdosage$  \\ %\midrule
\rowGrey \textbf{Parameter} & \textbf{Unit} & &  \textbf{Description} \\ %\midrule 
%\rowGrey Parameter & Unit & &  Description \\ \midrule 
$k_a$     &  $1/day$    &     & Absorption rate of lenograstim \\
$\mktr$   & $1/day$ &  & Transition rate       \\
$\mgamma$  & -- &  & Feedback speed of G-CSF on WBCs   \\
%Feedback exponent $\mbeta$  & -- \\
$\mslope$  & $L/\mu mol$ & &  PD effect of Ara-C on WBCs   \\
$\mbase$   & $10^9/L$  & &  Baseline of WBC count     \\
$x^0_{blasts}$     &  $10^9/L$  &  &  Relative number of blasts \\
              &            &  &   at start of consolidation therapy  \\ 
\rowGrey \textbf{State initial value } & \textbf{Value} & \textbf{State} & \textbf{Value} \\ %\midrule 
%\rowGrey Initial condition  & Value & &  \\ \midrule 
$x_{1},x_{2},{x}_{\text{exo1}},{x}_{\text{exo2}},x_D $ & 0& $x_g$ & $B_g$ \\
$\mxpr, \mxtr$ & $( \mbase \; \mkma )/\mktr$ & $x_{l1}$       &   \begin{tiny}$x^0_{blasts} ( \text{DB} \ \text{CR} ) - 0.005( \mbase \  \mkma )/\mktr$\end{tiny}  \\
$\mxma$        & $\mbase$ & $x_{l2}$       & \begin{tiny}$\mbase/99$ \end{tiny} \\
\bottomrule
\end{tabular}
}
%\medskip
\label{tab:ModelSpecifications}
\end{table*}

\subsection{Measurement functions}

The observed cell type measurements were WBC counts in the circulating blood and relative blast counts in the BM.
%The corresponding measurement functions $h(t)$ were formulated as%
The WBC count measurements were directly matched to the state $\mxma$ resulting in the corresponding measurement function
\begin{align}
h_{\text{wbc}}(t) & = \mxma(t).
\end{align}
The measurement function of the relative blast count was used from previous publications \cite{Pefani2013a, Pefani2014}:
\begin{align} \label{eq:MeasRelBlasts}
%h_{blasts}(t) &= 100 \ \frac{ x_{l1}(t)+0.005 \ \mxtr(t) }{ x_{l1}(t)+\mxpr(t)+\mxtr(t) + 300 }.
h_{blasts}(t) &= 100 \ \frac{ x_{l1}(t)+0.005 \ \mxtr(t) }{ \text{CR}_{ij} \ \text{DB} }
\end{align}
with the cellularity factor of patient $i$ in the $j$-th consolidation cycle
\begin{align}
\text{CR}_{ij} = \begin{cases} 0.2\phantom{0 } \quad if \ hypocellular \\ 0.4\phantom{0 } \quad if \ normocellular, \ years > 65  \\ 0.5 \phantom{0 } \quad if \ normocellular, \ years \le 65  \\ 0.95 \quad if \  hypercellular \end{cases}
\end{align}
and DB = $10^{12}$ being the approximated maximal tumor cell burden in acute leukemia \cite{Lichtman2006}.
%The assumed BM volume of approximately 1 liter, being in the range of published values \cite{Batinic1990, Nombela2017}, allows to specify both lineages in [$10^9 $ cells/L].
%We assume that the BM has a volume of approximately 1 liter being in the range of published values \cite{Batinic1990, Nombela2017} such that both lineages are determined in [cells$\times 10^9$/L].
%The numerator of the original function was extended with 0.5\% cells of the transit compartment.
As the measurement method for determining the relative blast counts in the BM did not differentiate between physiological and leukemic blasts the original function was extended with 0.5\% cells of the transit compartment. 
%, the numerator consists of the sum of the leukemic blast count $x_{l1}$ and 0.5\% cells of the transit compartment, which represents all stages of maturation between stem cells and matured cells.
Nombella and Manz \cite{Nombela2017} examined the range of the relative number of common myeloid progenitors in the BM to be 0.2-0.8\% represented in the function by 0.5\% cells of the transit compartment.

\subsection{Model predictions and treatment schedule optimization}

We analyzed the reliability of the newly developed model with out-of-sample cross validations. 
We thus predicted the last CC for all patients for whom measurements from more than one CC were available, based on models fitted to the measurements from all previous CCs.
%\todo{define personalized, individual model}
Additionally, we used the individual models of 14 patients for whom relative blast counts were available in the last CC for a mathematical optimization of the treatment schedules of the last CC.
We compared clinically important indicators such as nadir values and relative blast counts in the BM to the measured values.
%\todo{more detailed description of optimization procedure}
Optimizing the treatment schedule for patient $i \in \{1,\dots,20\}$ was formulated as a minimization problem
\begin{linenomath*}
\begin{subequations}\label{eq:OCProblem}
\begin{alignat}{2} 
	  %  \min_{x(t),u_C(t), u_{L}(t)  } &  x_{l1}(t_f) +  \int_{t_0}^{t_f} \frac{1}{\mxma(t)}dt + \int_{t_0}^{t_f} |u_C(t)|dt +\int_{t_0}^{t_f} |u_{L}(t)| dt  \label{LVFPeq:OCobjectiveFunc}         \\
            \min_{x^i(t),u^i_c(t), u^i_{l}(t)  } &  \text{OBJ}^i  \label{eq:OCobjectiveFunc}         \\
%\text{s.t.} \quad  & \text{Model }  \eqref{eq:ODEmodel}  \\
\text{s.t.} \quad  & \dot x^i(t) = f(x(t),\theta^i, u_c^i(t), u_l^i) \\
            &    x(t_0^i) = x_0^i  ,                                        \label{LVFPeq:x0}            \\
           % &    \mxma(t) > 1  ,                                        \label{LVFPeq:x0}            \\
           % &      \int_{t_0}^{t_f} u_C(t)  =                                \\ %       \label{LVFPeq:u}  
           % &\int_{t_0}^{t_f} u_{L}(t)  =                                \\ %       \label{LVFPeq:u}             
            &      u_c^i(t)  \in [0,2000]                               \\ %       \label{LVFPeq:u}  
            &      u_l^i(t)  \in [0,236]                                %       \label{LVFPeq:u}  
\end{alignat}
\end{subequations}
\end{linenomath*}
on the individual time horizon $t_0^i$ to $t_f^i$
with $f(\cdot)$ the mathematical model \eqref{eq:ODEmodel} from the previous section, $\theta^i = (\mbase^i, \mktr^i, \mslope^i, \mgamma^i, p_1^i)$ the empirical Bayes estimate resulting from the model fit to the measurements from all but the last CC,
$x^i_0$ the initial values of the ODE system at time point $t_0^i$,
$u_c^i(t)$ and $u_l^i(t)$ the control functions of Ara-C and lenograstim determining the administration schedule after optimization
and
\begin{align*}
\text{OBJ}^i =& \alpha_1 \; x^i_{l1}(t_f^i) +  \alpha_2 \; \int_{t_0^i}^{t_f^i} \frac{1}{{\mxma^i}^2(t)}dt \\ &+ \alpha_3 \; \int_{t_0^i}^{t_f^i} u_c^i(t)dt +\alpha_4 \; \int_{t_0^i}^{t_f^i} u_{l}^i(t) dt
\end{align*}
the objective function consisting of four terms.
The first term denotes the number of leukemic cells in the bone marrow at time $t_f^i$ representing the disease status at the end of the consolidation treatment.
The second term reflects the health condition of the patient during treatment (heavily penalizing small WBC counts). 
The last two terms model the costs via the amount of totally administered Ara-C and lenograstim, respectively. 
%Instead of defining the total amount of Ara-C and lenograstim as boundary constraints we defined them as objective function terms.
Scalar weights $\alpha_1,\dots,\alpha_4$ allow the weighting of these terms according to personalized, clinical, and ethical preferences.
Values of the weights were chosen by initial guesses and iteratively adapted until the clinically relevant optimization outcomes, meaning $h_{blasts}(t_f^i)< 5\%$ and $min(\mxma(t))> 1$, were met.
The final values for the $\alpha_i$ are presented in~\ref{tab:alphas} in the SM.
%Additionally, alternative formulations, e.g., with an upper bound (UB) on the number of leukemic blasts $x^i_{l1}(t) \le \text{UB}$ were considered.
%
%With these constants the two cost terms can have a minor influence compared to the clinically and ethically more important disease and immune system progression terms.

All optimization results were calculated for a time period starting 10 days before the start of the actual Ara-C treatment of the last CC ($t_0^i$) and ending with the time point of the patient's conducted BM puncture ($t_f^i$). 
This time horizon was chosen to compare the optimized values with the measured relative blast counts in the BM.
The initial conditions $x(t_0^i)$ were derived from the individual models.
We defined a hourly time grid for model evaluation and for Ara-C infusions in which Ara-C infusions can be optimized within the first 20 days. 
The control grid for Ara-C was restricted to the first 20 days so that no Ara-C infusions were placed at the end of the time horizon.
Lenograstim administrations were defined as 0.0007 day injections on the hourly grid once a day at 8 a.m. 
% and Ara-C infusions were not permitted on this interval.
The upper limit of hourly Ara-C infusions was chosen to be $2 \ g$ per hour, being the recommended maximum amount of a high-dose treatment schedule for a patient under 60 years with a BSA of $2 \ m^2$ which should not be exceeded \cite{Doehner2017}.
As mentioned in the introduction, intermediate Ara-C dosage is administered to elderly patients but in our previous study \cite{Jost2019} a 64 year old patient was treated with D12 such that we decided to define the upper limit of $2 \ g$ per hour as a further degree of freedom during optimization.
The upper limit of lenograstim administrations was chosen to be 263 $\mu g$ equivalently to the actual daily administered dose amount.
The infinite dimensional optimal control problem \eqref{eq:OCProblem} was solved by a direct collocation approach (simultaneous approach) in which the control functions and the differential states are simultaneously discretized by low order polynomials \cite{Biegler2010}.
The resulting finite optimization problem is large scale due to the introduction of additional optimization variables and constraints, but highly structured such that tailored iterative procedures can be applied to numerically calculate local optimal solutions.

\subsection{Model evaluation and software}

%%%% comment on IIV and residual variability
We aligned the nonlinear mixed-effects modeling to established PK/PD modeling approaches \cite{Friberg2002, Kloft2006, Quartino2014, Henrich2017}. % using similar PK/PD models und focusing on the same or similar biomarkers.
%We have oriented the nonlinear mixed-effect modeling based on several published modeling approaches using similar PK/PD models und focusing on the same or similar biomarkers.
Interindividual variability (IIV) was assumed to be log-normally distributed, and residual variability was estimated using an exponential error model.
%Other residual variability models were tested without improving model accuracy.
%An proportional error model was also tested with similar model accuracy but 
 %residual variability is modelled by a proportional error model.
%%%% comment on model evaluation and simulations
Model development was guided by objective function values, uncertainty of parameters, agreement of predicted and observed clinical end points and visual evaluation of the results through visual predictive checks with \textit{auto\_bin} option, goodness-of-fit plots, and (individual) weighted residuals over time.
%Log-likelihood ratio tests with $\alpha=0.001$ were used to evaluate the significant reduction of the final log-likelihood from the parameterized secondary PD effect model compared to the reference model \cite{Bertrand2008}.
Parameter estimation for the nonlinear mixed-effects modeling approach was performed using the first-order conditional estimation method with interaction algorithm implemented in NONMEM 7.4 (ICON Plc., Dublin, Irland).
%For the cross-validation analyzing Ara-C's secondary effects the first-order algorithm was used.
Standard errors were computed with the \$COVARIANCE step in NONMEM.
Pirana (Certara, Princeton, USA) was used for NONMEM execution and data analysis.
%The optimal control problems were formulated in CasADi \cite{Andersson2019} interfaced via Python2.7. %Collocation (radau), control grid, ipopt 
The optimal control problems were formulated in CasADi (Optimization in Engineering Center (OPTEC), K.U. Leuven) \cite{Andersson2019} interfaced via Python 2.7 (Python Software Foundation) by applying a simultaneous approach \cite{Biegler2007}.
The system of ordinary differential equations was discretized using direct collocation \cite{Biegler2007} with Lagrange polynomials with Legendre collocation points of order 3, and the nonlinear optimization problems were solved with Ipopt 3.12.3 \cite{Waechter2006}.

\section{Results}

\subsection{Modeling exogenous G-CSF}

\begin{figure*}[!t]
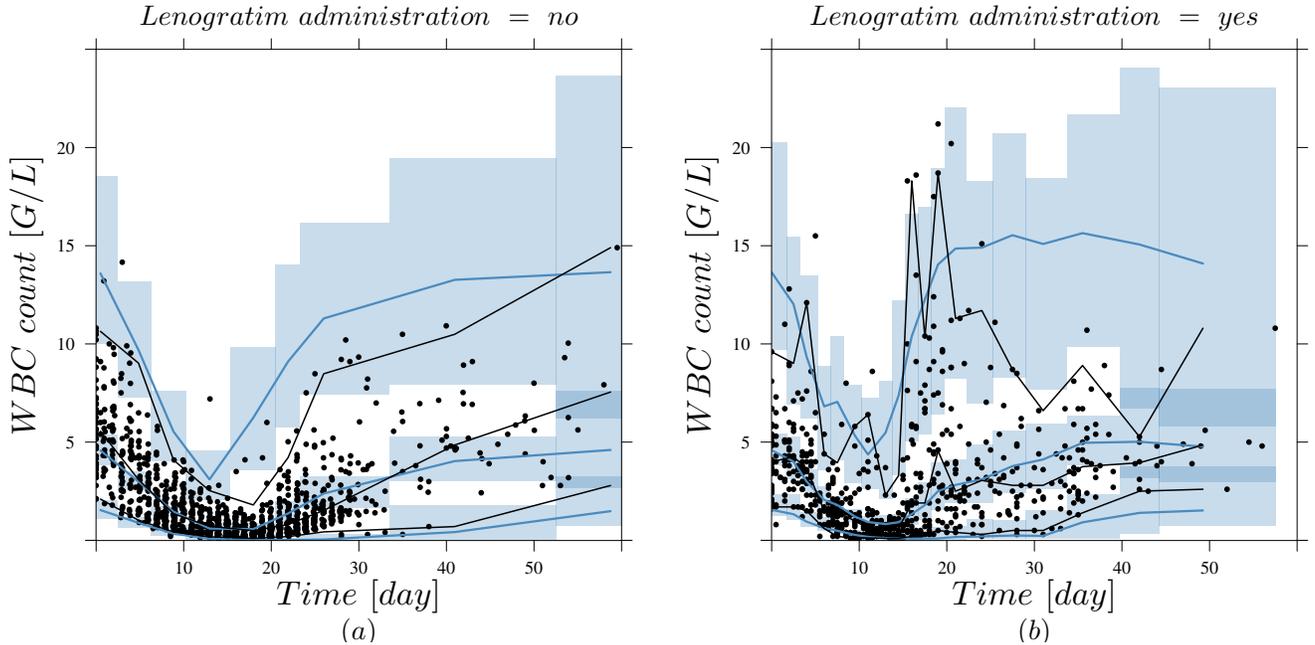

\begin{center}
\resizebox{1.0\textwidth}{!}{

}
\end{center} 
\caption{ 
Visual predictive checks (VPCs) stratified by lenograstim administration and derived by 1000 simulations with the final parameter estimates from the myelosuppression model with two transit compartments (fourth column of~\ref{tab:LenoModel}), for circulating WBCs (G/L) versus time (day).
Black dots are the measured WBC counts. 
Black and blue lines show the median and 2.5th and 97.5th percentiles of measurements and model predictions, respectively. 
The shaded areas represent the 95\% confidence intervals around the 2.5th, 50th and 97.5th percentiles of the model predictions. 
}
\label{fig:VPC_GoF_Cycles}
\end{figure*}

The effect of lenograstim on WBC counts was visually assessed by the cyclewise WBC dynamics after consolidation therapy.
Figure~\ref{fig:VPC_GoF_Cycles} shows the visual predictive checks derived from the final model after parameter estimation highlighting that patients who received lenograstim had a rapid increase of WBC counts during WBC recovery (compare the measurements in (a) and (b) after nadir).
Further, the figure underlines the good match between model and clinical data capturing the rapid increase of WBC counts during WBC recovery for patients receiving lenograstim.
The model slightly overpredicts the 50th and 97.5th percentiles for patients who did not receive lenograstim.
The evaluation of model fitting via the medians of the individual mean absolute errors and root mean squared errors in~\ref{tab:LenoModel} revealed that the extended myelosuppression model with a s.c. absorption model and two transit compartments described the clinical data best.
The first column in Table~\ref{tab:finalParameterEstimates} shows the estimated model parameters.
During model development we investigated the individual parameter estimates grouped by the lenograstim administration (\ref{fig:boxplotsLenograstim} in the SM).

\begin{table}[!t]
\centering
\caption{Results of parameter estimations for different PK/PD models and datasets. 
Shown are residuals (objective value), parameter estimates of fixed effects, interindividual variability as a coefficient of variation (CV\%) and exponential (exp.) residual errors as variance with relative standard errors (RSEs) in brackets
for in-sample and out-of-sample (without measurements of the last consolidation cycle).}
\begin{tabular}{|l|rrrr|}
%% column 1 /home/jost/mercurial/MATHOPT/PAPERS/191112_AML_AllCC_UlmAndMD_twoTransit
%% column 2  /home/jost/mercurial/MATHOPT/PAPERS/191111_AML_allDataNoBlasts
%% column 3  /home/jost/mercurial/MATHOPT/PAPERS/191111_AML_allData
%% column out-of-sample /home/jost/mercurial/MATHOPT/PAPERS/191111_AML_prediction 
\toprule      
\rowGrey           & \multicolumn{3}{c}{\textbf{in-sample}}                            & \textbf{out-of-sample} \\
          Dataset  & \textbf{cyclewise}  & \textbf{patientwise} & \textbf{patientwise}  & \textbf{patientwise}  \\ 
 Blasts   & \textbf{no}   & \textbf{no}    & \textbf{yes}   & \textbf{yes}  \\ 
\# patients        &     67           &        44         &    44    & 24 \\    %\hline 
 \rowGrey \multicolumn{5}{|c|}{\textbf{Fixed Effects (RSE\%)}}    \\
   $\mbase$        &     4.67(6)     &  4.67(7)         &      4.85(8)  & 4.50(11)\\    
   $\mktr$         &     0.196(10)    &  0.236(3)        &      0.218(6) & 0.224(7) \\   
   $\mslope$       &     10.1(3)     &  7.94(11)        &      8.53(8)  & 8.95(7)\\  
   $\gamma$        &     0.701(4)     &  0.651(4)        &      0.680(6) & 0.679(7)\\ 
   $k_a$           &     3.16(3)       &  3.15(26)        &      3.20(21) & 2.828(31)\\
   $x^0_{l1}$     &       --         &  --              &      0.029(28) & 0.0434(19)\\ 
\rowGrey \multicolumn{5}{|c|}{\textbf{Interindividual Variability CV\% (RSE\%) }}    \\  
   $\mbase$        &     39.6(8)     &  40.9(12)      & 49.3(14)    &      47.2(16)  \\  
    $\mktr$        &     19.3(10)     &  14.3(22)     & 25.2(21)    &      25.4(20)  \\   
    $\mslope$      &     42.1(13)     &  55.8(17)     & 11.8(24)    &      17.9(21)  \\  
    $\gamma$       &     19.8(10)     &  19.7(16)     & 21.2(15)    &      23.3(26)  \\ 
   $k_a$           &    103.3(16)     &  95.1(28)     & 58.6(34)    &      119.6(22)  \\
    $x^0_{blasts}$ &     --           &      --       & 67.9(23)    &      24.0(101)  \\
\rowGrey \multicolumn{5}{|c|}{\textbf{Residual Error (RSE\%)} }   \\ 
 Exp. error  &     0.152(9)    &  0.284(8)     & 0.315(8)    &      0.250(10)  \\
\bottomrule
\end{tabular}
\label{tab:finalParameterEstimates} 
\end{table}

\subsection{Modeling leukemic blasts}

The PK/PD model was fitted to the clinical data with and without consideration of the leukemic cell lineage.
The estimated parameter values are presented in the second and third column of Table~\ref{tab:finalParameterEstimates}.
The leukemic blast lineage only had a minor effect on the estimated parameter values with an increase of the slope parameter and the variance of the exponential error model. 
The model performance of describing the clinically observed circulating WBC counts and relative blast counts in the BM is shown as visual predictive checks (VPCs) in~\ref{fig:VPC_GoF} in the SM.
\ref{fig:GCSFsensitivity} shows the influence of G-CSF administrations (yes or no) and of varied G-CSF steady states on the WBC recovery.
%Ara-C without lenograstim administration resulted in a longer recovery time and a slightly lower WBC count before the start of the second and third CC.
%As a further consequence, the number of leukemic blasts in the BM was higher and increased more over time, than to the leukemic blast count when the actual treatment schedule of lenograstim was conducted. 
%A different G-CSF steady state affected the recovery time, where lower steady-state values provoke an overproduction of WBC counts, leading to a higher value than the WBC steady state.
%\begin{figure}[!t]
%\includegraphics[width=.49\textwidth]{AML190705_GcsfSteadyStateAndAdministration_148.eps} 
%\caption{WBC dynamics (solid black line) of the final model fitted to observed WBC counts of the first two consolidation cycles (blue dots), and the Ara-C and lenograstim treatment schedules of one exemplary patient are shown.
%The last cycle is used for model prediction and out-of-sample comparison. 
%Simulated WBC dynamics for no lenograstim (dotted black line) and for different G-CSF steady state values (from $20\%$ to $140\%$ of used value) are shown (solid gray lines). 
%No lenograstim administration prolonged WBC recovery time and lower/higher G-CSF steady state values shortened/prolonged WBC recovery.
%Moreover, no lenograstim administration resulted in a slightly larger leukemic blast count (the red area indicates the difference when compared to the actual treatment schedule shown in the third row). 
%}
%\label{fig:GCSFsensitivity}
%\end{figure}
We investigated the out-of-sample prediction performance of the final model with its extension to lenograstim and leukemic blasts and analyzed the potential of different treatment schedules derived by mathematical optimization.
Parameter estimation results for the data subset (compared to~\ref{fig:FlowChartStudies}) are shown in Table~\ref{tab:finalParameterEstimates}.
Compared to the in-sample parameter estimates, the values of $\mbase, \mktr, \mslope$ and $\mgamma$ are almost equal to the values derived from the whole dataset and the values of $k_a$ and $x^0_{l1}$ are slightly decreased, respectively increased.
The prediction performance is visualized as a goodness-of-fit plot in~\ref{fig:Prediction} in the SM.
%The blue circles are the measurements used for parameter estimation and the red cubes reflect the accuracy of the model predicting WBC counts in consecutive cycles.
Both in-sample and out-of-sample, the values are centered around the line of identity. No systematic error is apparent, only a slight overprediction of small WBC counts.
Finally, a global sensitivity analysis was performed to determine key parameters which contributed the most to variability in model outputs. 
The analysis can be found in the SM.

\subsection{Model predictions and optimal treatment schedules}

%\FJ{highlight daily administration of lenograstim}

%\FJ{highlight statement: our results indicate that two CCs with reduced 
%doses of Ara-C can achieve the same outcome as that achieved by one CC, with the benefit of 
%increased WBC nadir values}

\begin{figure*}[!t]
\begin{subfigure}[c]{0.49\textwidth}
\includegraphics[width=1.\textwidth]{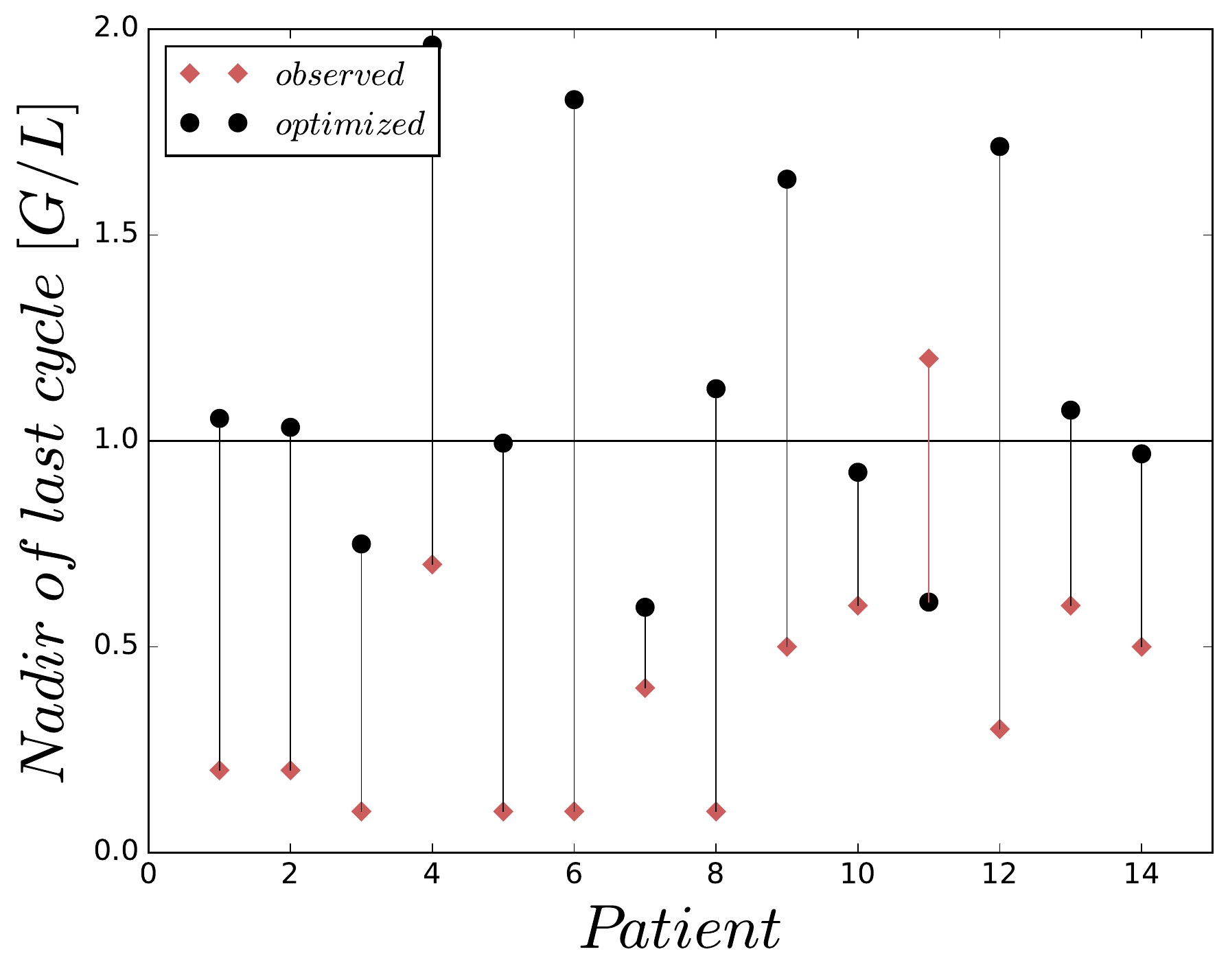} 
%\subcaption{ }
\end{subfigure}\begin{subfigure}[c]{0.49\textwidth}
\includegraphics[width=1.\textwidth]{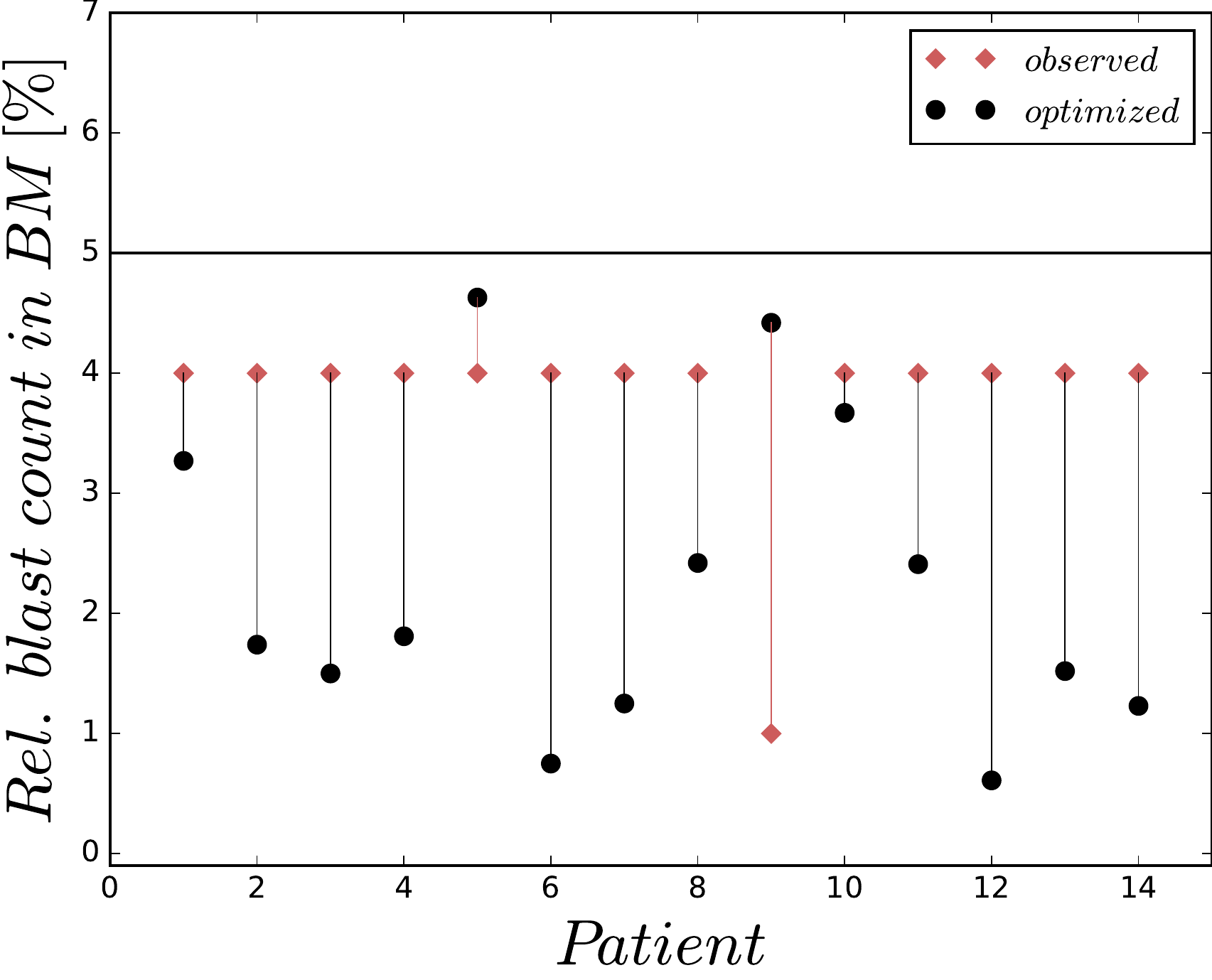}
%\subcaption{ }
\end{subfigure}
\caption{Comparison between clinical (observed) and optimized treatments with respect to white blood cell (WBC) count nadirs (left) and relative (rel.) blast counts in the bone marrow (BM, right) for 14 patients. The nadirs are significantly higher, often even above the leukopenia threshold, while the corresponding relative BM blasts are maintained in the same range as the observed values and below the clinically important threshold of $5\%$. }
\label{fig:OptimizedNadirsBlasts}
\end{figure*}

\begin{figure*}[!t]
\begin{center}
 \includegraphics[width=1.\textwidth]{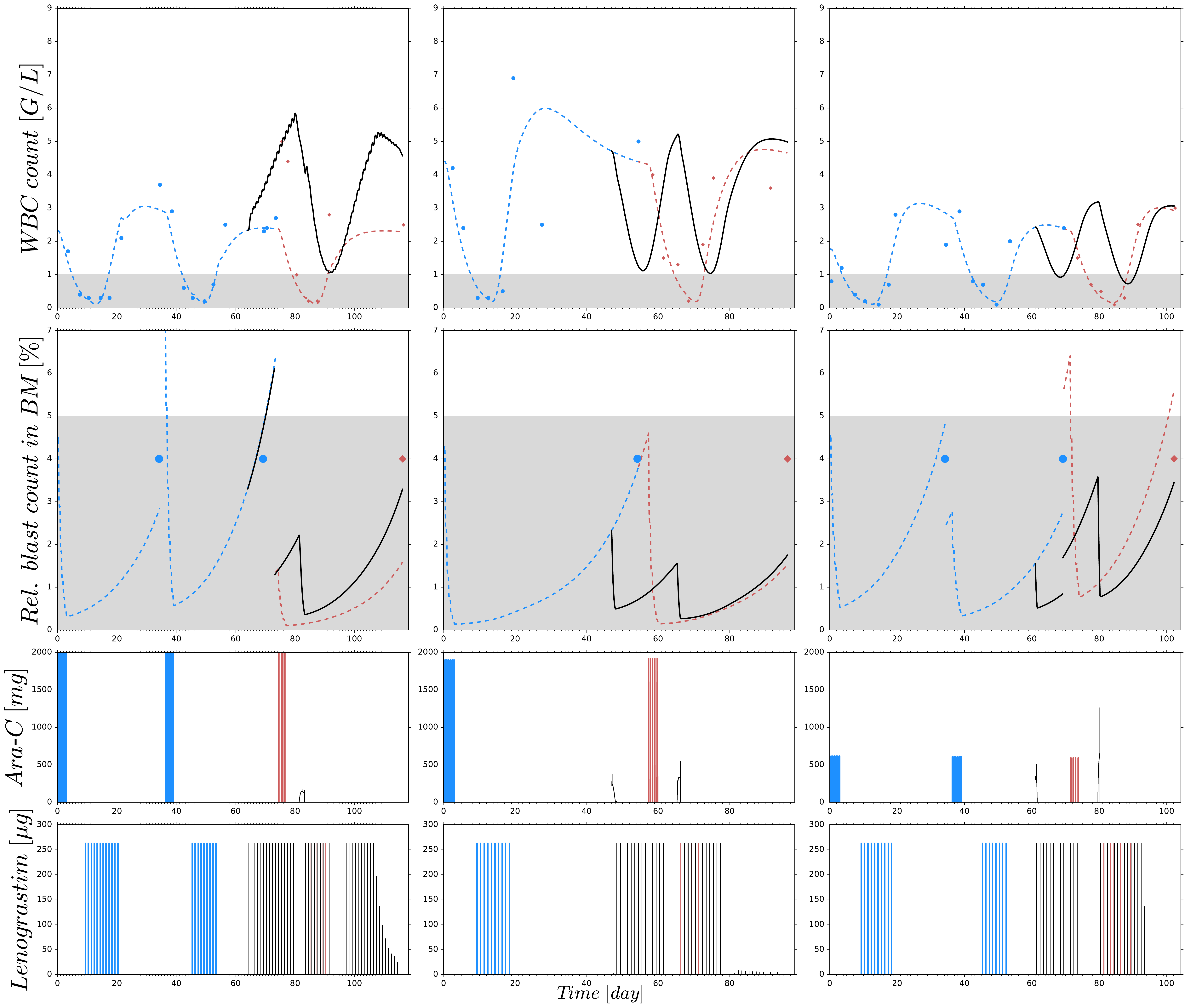} 
\end{center}
\caption{Detailed optimization results for Patients~1, 2, and 3, respectively (same order as in Figure~\ref{fig:OptimizedNadirsBlasts}).
The treatment schedules of Ara-C and lenograstim in the last CC were optimized (black) and compared with the clinically applied treatment schedules (red). 
Shown are WBC counts and relative blast counts in the bone marrow (BM, dotted lines) resulting from individual models. 
Personalization was performed using in-sample measurements (blue dots) and clinical treatment schedules (blue lines) from all but the last CC.
The optimized schedules and the affected dynamics of WBC (solid black lines) qualitatively differ for each patient. 
In (a), one later low-dose treatment and in (b), two intermediate-dose Ara-C treatments result in higher nadir values compared to the measured values (red dots). 
In (c), the daily lenograstim administrations before and after the postponed Ara-C treatment did not prevent a fall of WBCs below $1 \ G/L$. 
Discontinuities in the dynamics of the relative (rel.) blast counts in the bone marrow (BM) occur due to possible cyclewise cellularity changes in the measurement function \eqref{eq:MeasRelBlasts}.}
\label{fig:OCPsolution}
\end{figure*}

%\clearpage
Using the final model and the individual parameter estimates for 14 patients from above, we calculated optimized individual treatment schedules. 
Optimal refers to a numerical local optimization of \eqref{eq:OCProblem} in the last CC. 
From the solutions, we extracted the WBC nadir values and final time relative blast counts in the BM.
In Figure~\ref{fig:OptimizedNadirsBlasts}, a comparison between the optimized and clinically observed values shows, that the optimized treatment schedules of Ara-C and lenograstim achieved an increase in nadir values for each patient (in median 4.2-fold higher values), although relative BM blast counts were comparable to the observed ones and below the clinically important threshold of $5\%$. 
Not shown is that the median Ara-C amount was lower by approximately 60\%.
%For eight patients the optimized treatment schedules achieve nadir values above the $1 \ G/L$ threshold by either a later Ara-C treatment (compare Figure~\ref{fig:OCPsolution}a) or two low-dose treatments resulting in two CCs instead of the actual one CC (compare Figure~\ref{fig:OCPsolution}b).
Three exemplary optimization results with detailed trajectories are shown in Figure~\ref{fig:OCPsolution}, and the results for all 14 patients are shown in the SM.
For all patients the optimal treatment schedules suggested lenograstim administrations almost every day  (except during the Ara-C infusions) accelerating the WBC recovery to its baseline value and raising the WBC count before Ara-C treatment.
%Important to mention is the observation, that within the optimal treatment schedules lenograstim should be administered almost every day (expect for the Ara-C infusions).
While optimal timing and dosages of Ara-C and lenograstim were individually optimized and hence different for each considered patient, two qualitative patterns could be observed. 
In pattern A, an additional Ara-C administration period (and hence an additional CC) was introduced, and the administration order was Ara-C, lenograstim, Ara-C, lenograstim. 
In pattern B, the nadir was increased compared to the clinical treatment schedule with the administration order lenograstim, Ara-C, lenograstim. 
The amount of Ara-C was usually considerably reduced. 
Figure~\ref{fig:OCPsolution} shows examples for patterns A (middle, right) and B (left) with in total 9 times pattern A and 5 times pattern B.
Concluding, the optimal treatment schedules resulted in increased WBC nadir values simultaneously achieving similar relative leukemic cell counts in the BM via the reduction of Ara-C dosage and the introduction of additional CCs.

\section{Discussion} %

The development, fitting, validation, and analysis of the PK/PD model was performed in an iterative way starting with the modification of the myelosuppression model provided by Quartino and colleagues \cite{Quartino2014} to Ara-C and to the s.c. administration of lenograstim and completed with the incorporation of the leukemic blast lineage.
Several parameter estimations were performed to fit and validate the models.
An analysis of the estimated model parameters in Table~\ref{tab:finalParameterEstimates} showed that the fixed effects and interindividual variabilities were in the same ranges in all the numerical studies, indicating that the general model behavior was maintained despite model extensions.
Equivalent to the results in \cite{Quartino2014}, our model predictions revealed an overprediction of the 50th and 97.5th percentiles for patients who did not receive lenograstim (c.f. Figure~\ref{fig:VPC_GoF_Cycles}).
During model development, we examined two other models \cite{Jayachandran2014,Stiehl2018}. 
Both models also showed discrepancies in predicting the 97.5th percentile for the model from \cite{Jayachandran2014} and the 50th percentile for the model from \cite{Stiehl2018}.
This line of research was not further investigated.
The values of $\mbase$ were within the normal human WBC range of 4-10 $G/L$ and coincided with the neutrophil base value from Quartino \textit{et al.} \cite{Quartino2014}, assuming that the relative amount of neutrophils ranges between 60-70$\%$.
Compared to published WBC baseline values for the model by Friberg \textit{et al.} ranging between $7$ and $7.8 \ G/L$ \cite{Friberg2002}, our values were 2-3 $G/L$ lower.
The mean maturation times of 102-122 hours are reasonable and fit into the range of previously published values \cite{Evans2018}.
The value of $\mgamma$ was larger compared to the model of Quartino \textit{et al.} \cite{Quartino2014}, which might be due to the dense treatment schedules. 
The residual error doubled from the cyclewise to patientwise management of the data, assuming that interoccasional variabilities, which were not the focus of this work and as a consequence not modeled, might be one of the reasons for an increased model-reality mismatch. 
This mismatch was further increased with the consideration of relative blast counts in the third column of Table~\ref{tab:finalParameterEstimates} introducing an additional source of error.

We visualized the parameter estimates separately for cycles in which lenograstim was administered or not to analyze the influence of lenograstim on the parameter estimates.
Figure~\ref{fig:boxplotsLenograstim} in the SM shows that the steady state value of WBC is lower for patients receiving lenograstim, indicating that the demand of exogenous G-CSF might be related to the patients' WBC steady states.
In comparison to the estimated values of $\mgamma$ which were almost equal between the two groups after modeling the lenograstim administration, the transport rate $\mktr$ was still increased for the patients who received lenograstim although exogenous G-CSF already influenced proliferation and maturation via the feedback term $({x_g}/{B_g})$.
We suspect that the higher values agglomerate biological phenomena that are not correctly described or fully covered by the current model.
Nevertheless, the model exhibited the same behavior as in clinical trials with pegfilgrastim \cite{Jaramillo2017}, i.e., a prolonged WBC recovery time of several days without G-CSF, compared to~\ref{fig:GCSFsensitivity}. 

The model has several constant parameters that were fixed to published values. As only WBC and relative BM blast counts were observed, this was necessary to avoid overfitting and obtain a good predictive accuracy. However, the interpretation of parameter values could now be misleading, as incorrect constants and modeling are usually compensated by parameter values.
A better data situation with additional G-CSF and Ara-C concentration measurements would allow to identify further parameters.
In addition to the global sensitivity analysis, in which we investigated the key parameters having the largest impact on two clinically relevant model outputs, a structural sensitivity analysis (\cite{Gutenkunst2007,Chics2011,Villaverde2019}) would help to systematically investigate identifiability of parameters assuming additional biomarker measurements.
%\FJ{comment on bioavailability and sensitivity analysis}.

The VPC in~\ref{fig:VPC_GoF} revealed that the model of leukemic blasts is able to describe trends respectively overpredicts the measurements with its exponential behavior, leading to a purely leukemic steady state after 4.6 month being in the reported interval for remission before relapses occurred \cite{Cassileth1988}.
%This modeling assumption is very conservative as we assume that at some future timepoint the leukemic blasts will rise controlling and suppressing the healthy WBCs.
This conservative model behavior was chosen to study the impact of different treatment schedules on the increase of leukemic cells and might not exactly represent the patients actual leukemic blast dynamics. 
Therefore, the presented treatment optimization results have to be considered with care and further investigations and efforts have to be undertaken to develop more advanced and reliable models for bringing optimized treatment schedules to clinics.
As the number of leukemic blasts in our model will eventually converge to a purely leukemic steady state, we can only compare short-term impacts of treatment schedules on leukemic blasts and hence relapse probabilities. 
In the future, additional modeling assumptions could be considered, e.g., stable steady states of coexistence between leukemic and healthy cells achieved via the inclusion of the leukemic blasts' steady state value in the zero-order production term of endogenous G-CSF or a threshold value of leukemic blasts below which the immune system could avoid a relapse for good.
Modeling minimal residual disease (MRD), proposed by multiple recent studies as a strong prognostic marker for relapse in AML \cite{Zhu2013,Rucker2019,Thol2018,Schuurhuis2018} might also be a promising alternative to leukemic blasts.
In the current study, no MRD information was available such that we concentrated on a model describing the relative blast count measurements.
In the current study we focused on dynamic deterministic models but the low number of BM measurements might force future model development to stochastic or survival analysis approaches as it was previously done by \cite{Dinh2019} in their proposed stochastic MRD model.

% Secondary pharmacodynamic effect of Ara-C
In our previous model \cite{Jost2019}, secondary PD effect of Ara-C were analyzed and an empirical model extension through a second PD effect on the feedback term $\mgamma$ was proposed.
During model development, we tested a parameterized PD effect. However, the evaluation criteria (such as the root mean squared error or a cross-validation in which the model was fitted to standard schedules and validated on dense schedules) showed only a minor benefit resulting from the consideration of Ara-C's possible secondary effects.
For this reason and without any concentration-time profiles of Ara-C we decided to neglect a secondary PD effect of Ara-C.
An additional simplification was made.
We used the plasma concentration of Ara-C for the PD effect although the side of action is within the bone marrow and it was shown that Ara-C plasma concentration is not the best predictor for bone marrow and intracellular Ara-C activity \cite{Rustum1982,Liliemark1985}.
To obtain a physiologically-based PK and PD model of Ara-C including secondary effects, further studies have to be performed to analyze the mechanisms and metabolism of high-dosage Ara-C \cite{Li2017} and its impact on dense treatment schedules.
%First efforts in physiologically-based PK model of Ara-C were presented by Dedrick et al. in 1973 and serves as a starting point \cite{Dedrick1973}.
A first physiologically-based PK model was already presented in 1973 by Dedrick et al. and serves as a starting point \cite{Dedrick1973}.

%It is a long way until the presented optimized individual treatments schedules from optimization problems would be a realizable alternative to current schedules.
%We want to discuss several arising drawbacks from our case study.

The optimized individual treatments derived by solving problem~\eqref{eq:OCProblem} rely on the mathematical model \eqref{eq:ODEmodel}. 
Application of the results to the real world is thus always under the assumption that the model and a personalized parameter estimation capture reality sufficiently well. 
This model-reality mismatch is amplified when optimized results are calculated. It is well known that optimization tends to exploit modeling errors as the ones discussed above. Thus, all interpretations should be considered very carefully and should be mainly seen as an incentive for clinical trials to validate the conjectures derived from simulations.
On the positive side, the developed model showed a good prediction accuracy for a variety of different treatment schedules despite the large number of constants and missing concentration measurements of Ara-C and G-CSF.
This could not only become a basis for individual decision support, but allowed for the first time to quantify the potential of optimized treatment schedules in terms of nadir values, blast counts, and overall chemotherapy usage. We see the value of a more than 4-times increased nadir as a strong motivation to continue research in model-based treatment planning, even if the current personalized models might not yet be a perfect match to the situation of the patient for whom the data were observed.
Additionally, the approach allows to apply a variety of methods from mathematical optimization to get closer to clinical practice. 
Our deterministic optimization approach is very sensitive with respect to model parameters and the choice of scalar weights within the objective function.
Stochastic optimization techniques result in optimized schedules that are more robust against modeling and parameter uncertainties \cite{Ben-Tal2006}. The consideration of combinatorial constraints restricting the administration of Ara-C and lenograstim to plausible schedules would increase the applicability of the optimized schedules in clinical practice. % \cite{Gerdts2012,Sager2015}. 
Multi-objective optimization can provide Pareto fronts with respect to key performance indicators (high WBC, low blasts, low costs, low treatment time, \ldots) as already indicated in this study. % comment
The optimized treatment schedules demonstrated that a  60$\%$ (median) reduction in the amount of Ara-C and daily administrations of lenograstim could lead to higher nadir values compared to the clinical schedules (see Figure~\ref{fig:OptimizedNadirsBlasts}a).
The efficacy of the optimized treatment schedules was evaluated by comparing the optimized and measured relative blast counts in the BM at the end of the last CCs (c.f. Figure~\ref{fig:OptimizedNadirsBlasts}b).
The first clinical impact of the exploration of the optimized treatment schedules was the proposed administration of lenograstim before the start of Ara-C treatments, similar to the FLAG protocol \cite{Montillo1998}, as a prevention to mitigate myelosuppression and increase leukemic blast death.
For all 14 patients, lenograstim accomplished an increase in WBC count before Ara-C treatment, leading to moderate myelosuppression compared to the conducted treatment schedules (see Figure~\ref{fig:OptimizedNadirsBlasts}).
In the clinical trial from which the dataset was provided \cite{Schlenk2019}, lenograstim administration was started nine days after Ara-C treatment, reducing the WBC recovery time but not necessarily achieving nadir values above 1 $G/L$.
%The daily lenograstim administrations provoke an increased WBC count value alleviating the Ara-C induced myelosuppression.
As we considered the amount of lenograstim within the objective function, we assessed the times that had the smallest or largest impact either on WBC recovery or leukemic blast apoptosis.
We also performed calculations with a modified objective function without consideration of WBC count and lenograstim costs ($\alpha_2=\alpha_4=0$). The optimized treatment schedules still resulted in the administration of lenograstim every day. This indicates that exogenous G-CSF has a beneficial influence on the eradication of leukemic blasts.  %\todo{perform computation without cost of G-CSF and healthiness -> Result: always give GCSF}
%This known fact \cite{Poston2017} is shown in Figure~\ref{fig:GCSFsensitivity}.
%The influence of G-CSF on the leukemic cells is a  \cite{Poston2017}
%It is known that G-CSF also has an influence on the leukemic cells \cite{Poston2017} and this is shown in Figure~\ref{fig:GCSFsensitivity}.
In our model and setting, lenograstim administration not only reduced WBC recovery times but also the leukemic blast counts (c.f.~\ref{fig:GCSFsensitivity}). This coincides with clinical findings \cite{Nomdedeu2015}.
However, the contrary assumption also exists: exogenous G-CSF may lead to an increased leukemic blast count.
Until now, no evidence was given which claim holds, and in general, no clinical trial with long-term follow-up has shown an increase in mortality or relapse rate if G-CSF was administered \cite{Ohno1993,Poston2017}.
As the optimized treatment schedules propose daily administration of lenograstim, the change from s.c. injections to continuous intravenous infusions might be worth considering.
However, it was shown that the s.c. administration of G-CSF (filgrastim) results in lower peaks but more prolonged and stable levels of G-CSF compared with intravenous administration \cite{Paul2014}.
Considering short-term effects under the assumption of rapidly evolving leukemic blasts, our results indicate that two CCs with reduced doses of Ara-C can achieve the same outcome as that achieved by one CC, with the benefit of increased WBC nadir values. This pattern emerged in 9 out of 12 cases and coincides with published results for docetaxel-induced neutropenia \cite{Patel2014}. This result gives a partial answer to the question of Schlenk regarding whether four cycles of consolidation therapy are the best treatment choice \cite{Schlenk2016}. 
The developed mathematical model and optimization approach might help in the future to determine an optimal treatment schedule for the whole consolidation phase.

\section{Conclusion}

We developed a PK/PD model for the consolidation phase of AML patients treated with Ara-C and lenograstim.
The model was able to predict the dynamics of WBCs in consecutive cycles and helped to understand the interaction between WBCs and leukemic blasts and how they respond to different treatment schedules.
The developed model and the results from the computational approach to optimize the administration of Ara-C and lenograstim with respect to clinically important outcomes are further steps toward providing personalized medicine and decision-support tools for physicians \cite{Sager2018}.
%Further studies with more observed biomarkers need to be performed for a better model calibration.
Although the mathematical model might not capture all relevant processes accurately and a direct transfer of individually optimized schedules into clinical practice is not recommended at this stage, out results give for the first time a quantification of the potential of mathematically optimized AML consolidation treatment.
The more than 4-fold higher WBC counts at nadir at comparable simulated relative bone marrow blast counts are encouraging to pursue this line of research.

% or


%\appendix 

\bibliographystyle{IEEEtran}
\bibliography{ieeeBiomedicalEng}

\clearpage
\newpage

\setcounter{page}{1}
\section*{supplemental material}

\renewcommand{\thefigure}{Figure S\arabic{figure}}    
\renewcommand{\figurename}{}
\setcounter{figure}{0}
\renewcommand\thetable{Table S\arabic{table}}    
\renewcommand{\tablename}{}
\setcounter{table}{0}

\begin{figure*}[!h]
\begin{center}
\resizebox{.49\textwidth}{!}{
\includegraphics[width=0.8\textwidth]{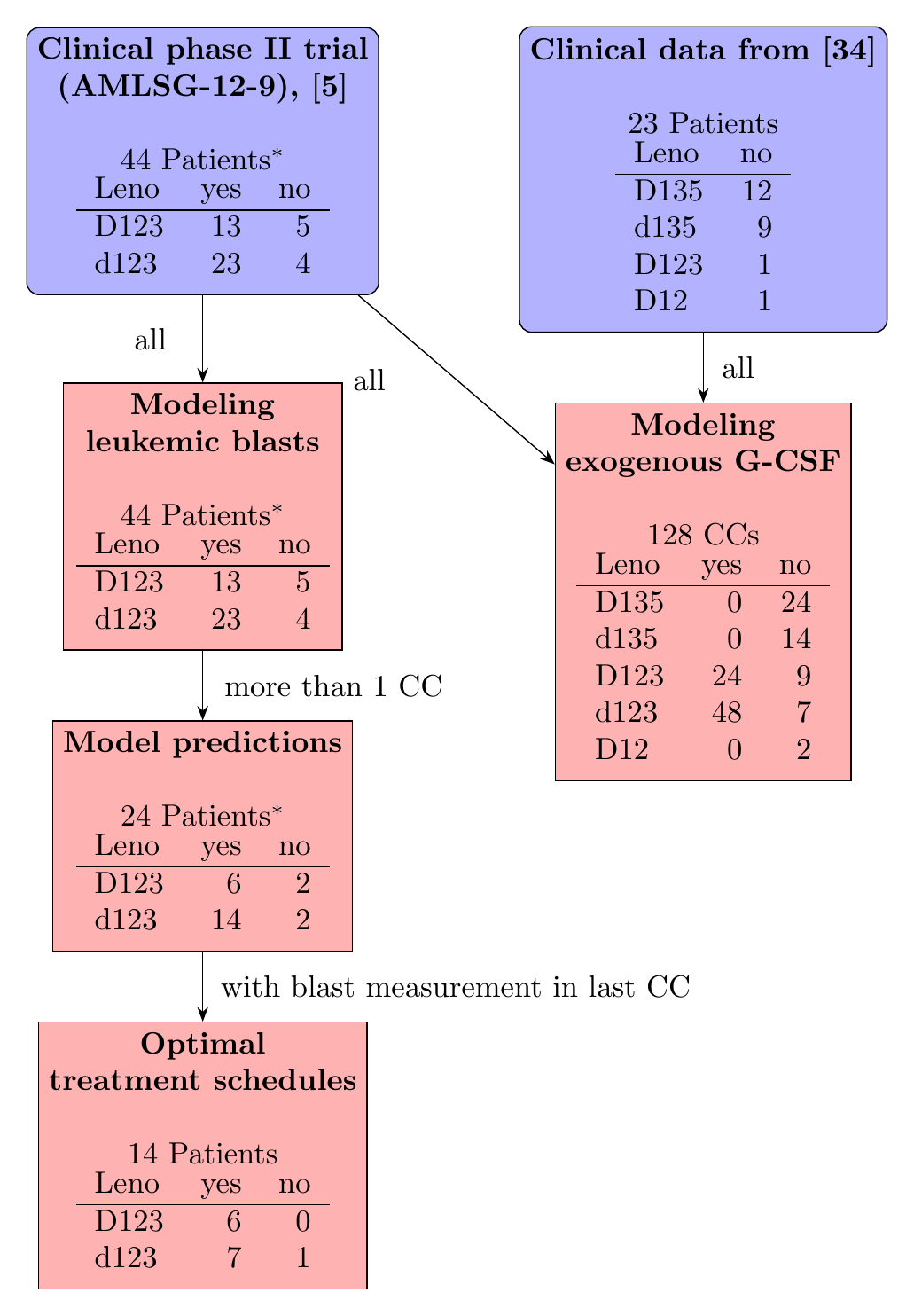}
}
\end{center}
\caption{Diagram of the two datasets and their subsets used in the different sections and for the pharmacodynamic (PD) modeling. Choices of subsets were based upon data availability, e.g., administration of lenograstim (Leno), granulocyte-colony stimulating factor (G-CSF), and numbers of consolidation cycles (CCs). One patient $^*$ received 1 CC without and 2 CCs with Leno, the data were split.}
\label{fig:FlowChartStudies}
\end{figure*}

\begin{table*}[!h]
\caption{Model constants, patient-specific constants, model parameters, and initial values with their units and descriptions.
%The constants of the PK model for Ara-C were determined from published data \cite{Kern1997, Jost2019} and applied to all patients.
} 
\label{tab_Constants}
\centering \footnotesize
\bgroup
\def\arraystretch{.9}
\setlength\tabcolsep{3.5pt}
%\resizebox{1.\textwidth}{!}{
\begin{tabular}{|lcrl| }
\toprule
\rowGrey \textbf{Constant} & \textbf{Unit} & \textbf{Value} & \textbf{Description}\\
%  Constant & Unit & Value & Description\\
%\midrule
  \multicolumn{4}{|l|}{ \textbf{PK model of Ara-C}} \\ 
$\mkaz$          &  $1/day$  & $98.2920$  & Elimination rate of Ara-C from \cite{Jost2019} \\
$\mkab$          &  $1/day$  & $2.6616$   & Distribution rate of Ara-C from \cite{Jost2019}  \\
$\mkba$          &  $1/day$  & $12.8784$  & Distribution rate of Ara-C from \cite{Jost2019}  \\
$\mV$                &  $L$  & $37.33$ & Volume of central compartment from \cite{Jost2019}  \\ % $37.33$ \\
$\mMM$       &  $g/mol$  & $243.217$  & Molecular mass of Ara-C \\
$dur_c$       &  $day$  & $1/8$  & Infusion time \\
  \multicolumn{4}{|l|}{\textbf{PD model of WBCs and leukemic blasts}} \\ 
$\mkma$            &  $1/day$    & $2.3765$ & Death rate of circulating WBCs from \cite{Quartino2014} \\
$B_g$              &   $ng/L$    &     24.4   & Endogenous G-CSF steady state from \cite{Quartino2014}   \\
$k_{e,g}$            &  $1/day$    &    0.592$\times$24  & Non-specific elimination rate constant from \cite{Quartino2014} \\
$k_{ANC}$          &  $1/day$    &    5.64$\times$24 & Neutrophil-dependent elimination rate from \cite{Quartino2014} \\
$\beta$            &  -          & 0.234  &  Feedback regulation of G-CSF on WBCs \\
$a_1$              &     -       & 0.875  &  Probability of self-renewal from \cite{Stiehl2018}  \\
$p_1$              &     $1/day$ & 0.1  &  Leukemic cell proliferation rate from \cite{Stiehl2018}  \\
$d_2$              &     $1/day$ & 2.3  &  Leukemic cell death rate from \cite{Stiehl2018} \\
$c_1$              &   $L/10^9$  &  0.01   &   G-CSF quasi steady-state feedback scaling factor \\ % constant     \\
$c_2$              &   $L/10^9$  &  0.01  &   G-CSF quasi steady-state feedback scaling factor \\ % constant      \\
  \multicolumn{4}{|l|}{\textbf{PK model of lenograstim}} \\ 
$k_{a2}$           &  -    & $\frac{10}{3}k_{a1}$  & 2. absorption rate of lenograstim \\
$V_g$                       &  $L$         &  14.5 & Volume of distribution from \cite{Hayashi1999}  \\ % \midrule
$dur_l$       &  $day$  & $0.0007$  & Infusion time \\
%\midrule \rowGrey Patient-Specific &  Unit  & Range  & \\ \midrule
$\mBSA$            &  $m^2$  & $[1.61,2.07]$ & Body surface area (patient-specific) \\
$u_{c}(t)$             &  $g/m^2$ & $[1,3]$  & Ara-C $\mdosage$ (patient-specific) \\
$u_{l}(t) $     &  $\mu g$ & $\{263,324\}$  & Lenograstim $\mdosage$ (patient-specific)  \\ %\midrule
\rowGrey \textbf{Parameter} & \textbf{Unit} & &  \textbf{Description} \\ %\midrule 
%\rowGrey Parameter & Unit & &  Description \\ \midrule 
$k_{a1}$     &  $1/day$    &     & 1. absorption rate of lenograstim \\
$\mktr$   & $1/day$ &  & Transition rate       \\
$\mgamma$  & -- &  & Feedback regulation of G-CSF on WBC proliferation   \\
%Feedback exponent $\mbeta$  & -- \\
$\mslope$  & $L/\mu mol$ & &  PD effect of Ara-C on WBCs   \\
$\mbase$   & $10^9/L$  & &  Baseline of WBC count     \\
$x^0_{blasts}$     &  $10^9/L$  &  &  Relative number of blasts \\
              &            &  &   at start of consolidation therapy  \\ 
\rowGrey \textbf{State initial value } & \textbf{Value} & \textbf{State} & \textbf{Value} \\ %\midrule 
%\rowGrey Initial condition  & Value & &  \\ \midrule 
$x_{1},x_{2},{x}_{\text{exo1}},{x}_{\text{exo2}},x_D $ & 0& $x_g$ & $B_g$ \\
$\mxpr, \mxtr$ & $( \mbase \; \mkma )/\mktr$ & $x_{l1}$       &   \begin{tiny}$x^0_{blasts} ( \text{DB} \ \text{CR} ) - 0.005( \mbase \  \mkma )/\mktr$\end{tiny}  \\
$\mxma$        & $\mbase$ & $x_{l2}$       & \begin{tiny}$\mbase/99$ \end{tiny} \\
\bottomrule
\end{tabular}
}
%\medskip
\label{tab:ModelSpecifications}
\end{table*}

\begin{table*}[!h]
\caption{Medians of individual mean absolute errors (MAE) and root mean squared errors (RMSE) with standard deviations in parenthesis for different myelosuppression models with and without consideration of lenograstim. The consideration of lenograstim (Leno) described via a single pathway absorption model iteratively increases the model fits by the inclusion of additional transit compartments (Transit) until the best fit was achieved with one transit compartments.
%The constants of the PK model for Ara-C were determined from published data \cite{Kern1997, Jost2019} and applied to all patients.
} 
\label{tab:LenoModel}
%\resizebox{.49\textwidth}{!}{
\begin{center}

}
\subcaption{VPC of WBC counts.}
\end{subfigure}
\begin{subfigure}[c]{0.49\textwidth}
\resizebox{1.05\textwidth}{!}{
%/home/jost/nonmem_EXAMPLES/pirana/examples/191111_AML_allData/vpc_ModelGammaLeno_191110_onlyBlasts/PsN_vpc_plots.tex
%
}
\subcaption{VPC of relative blast count in the BM.}
\end{subfigure}
\caption{Visual predictive checks (VPCs), derived by 1000 simulations, for
circulating WBCs [G/L] and relative (rel.) blast counts in the bone marrow (BM) [$\%$] versus time [day]. Black dots are the measured WBC counts, respectively rel. blast counts in the BM. 
Black and blue lines show the median and 2.5th and 97.5th percentiles of measurements and model predictions, respectively. 
The shaded areas represent the 95\% confidence intervals around the 2.5th, 50th and 97.5th percentiles of the model predictions. }
\label{fig:VPC_GoF}
\end{figure*}

\begin{table*}[!h]
\centering
\caption{Final values of the multiobjective optimization problem weights.}
\begin{tabular}{l|rrrr}
       Patient & $\alpha_1$ & $\alpha_2$ & $\alpha_3$ & $\alpha_4$ \\ \hline
             1 &    .9 & 3.  &  0.001 & 0.007    \\
             2 &  .8   & 1.  & 0.0001 & 0.007    \\
             3 &  1.8  & .8  & 0.0001 & 0.007    \\
             4 &  1.   & 2.  & 0.001  & 0.007   \\
             5 &  .35  & 1.  & 0.0001 & 0.007    \\
             6 &  1.   & 1.  & 0.0001 & 0.007    \\
             7 &  1.4  & 1.  & 0.0001 & 0.007    \\
             8 &  1.   & 1.  & 0.0001 & 0.007    \\
             9 &  1.   & 10. & 0.0001 & 0.007    \\
             10&  1.25 & 1.  & 0.001  & 0.007    \\
             11&  1.   & 1.  & 0.001  & 0.007    \\
             12&  .6   & 1.  & 0.0001 & 0.007    \\
             13&  1.   & 1.  & 0.0001 & 0.007    \\
             14&  .9   & 1.  & 0.001  & 0.007    
\end{tabular}
\label{tab:alphas}
\end{table*}

\begin{figure*}[!t]
\begin{subfigure}[c]{0.49\textwidth}
\begin{center}
\includegraphics[width=1.\textwidth]{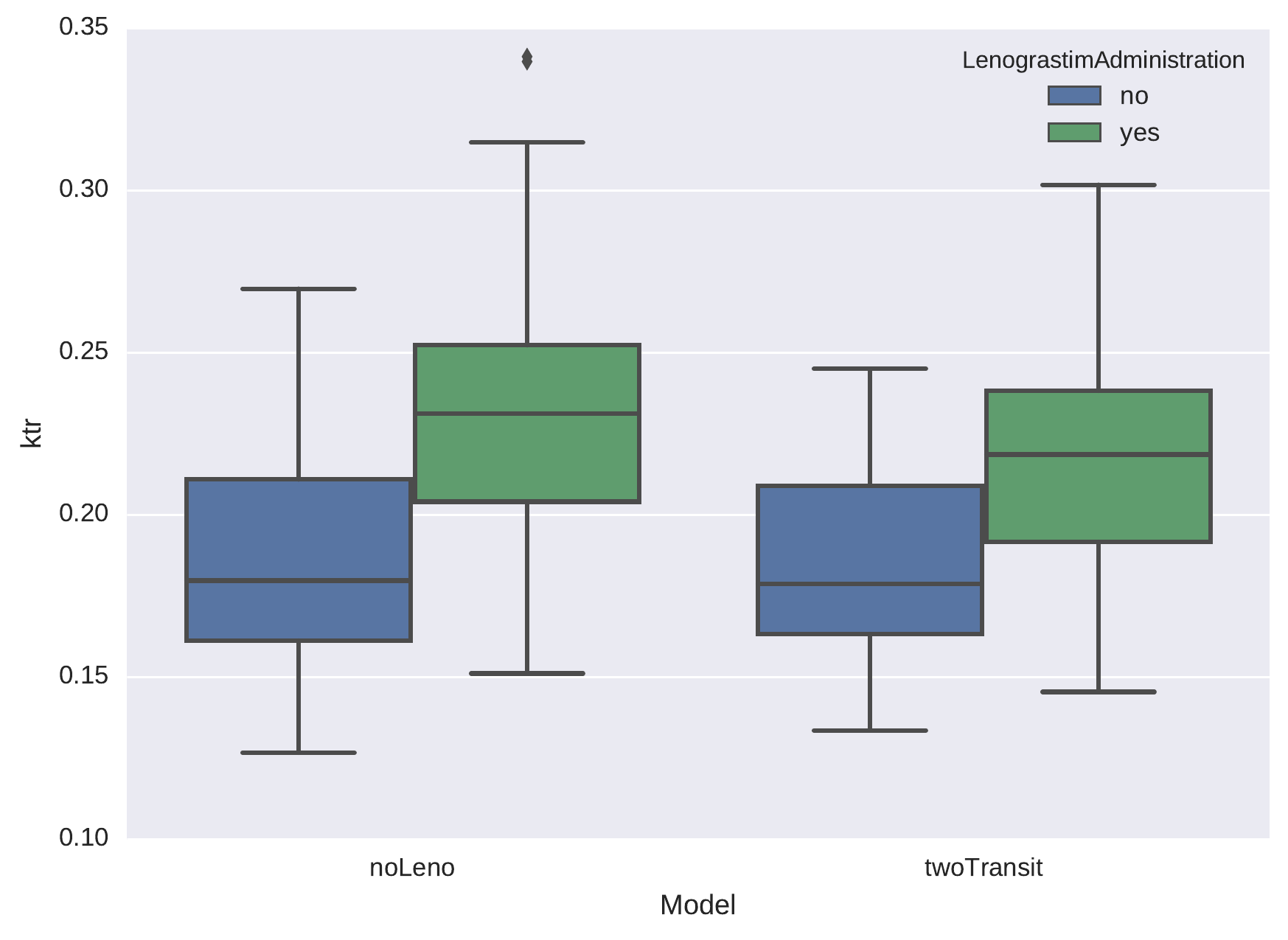} 
%\subcaption{}
\end{center}
\end{subfigure} 
\begin{subfigure}[c]{0.49\textwidth}
\begin{center}
\includegraphics[width=1.\textwidth]{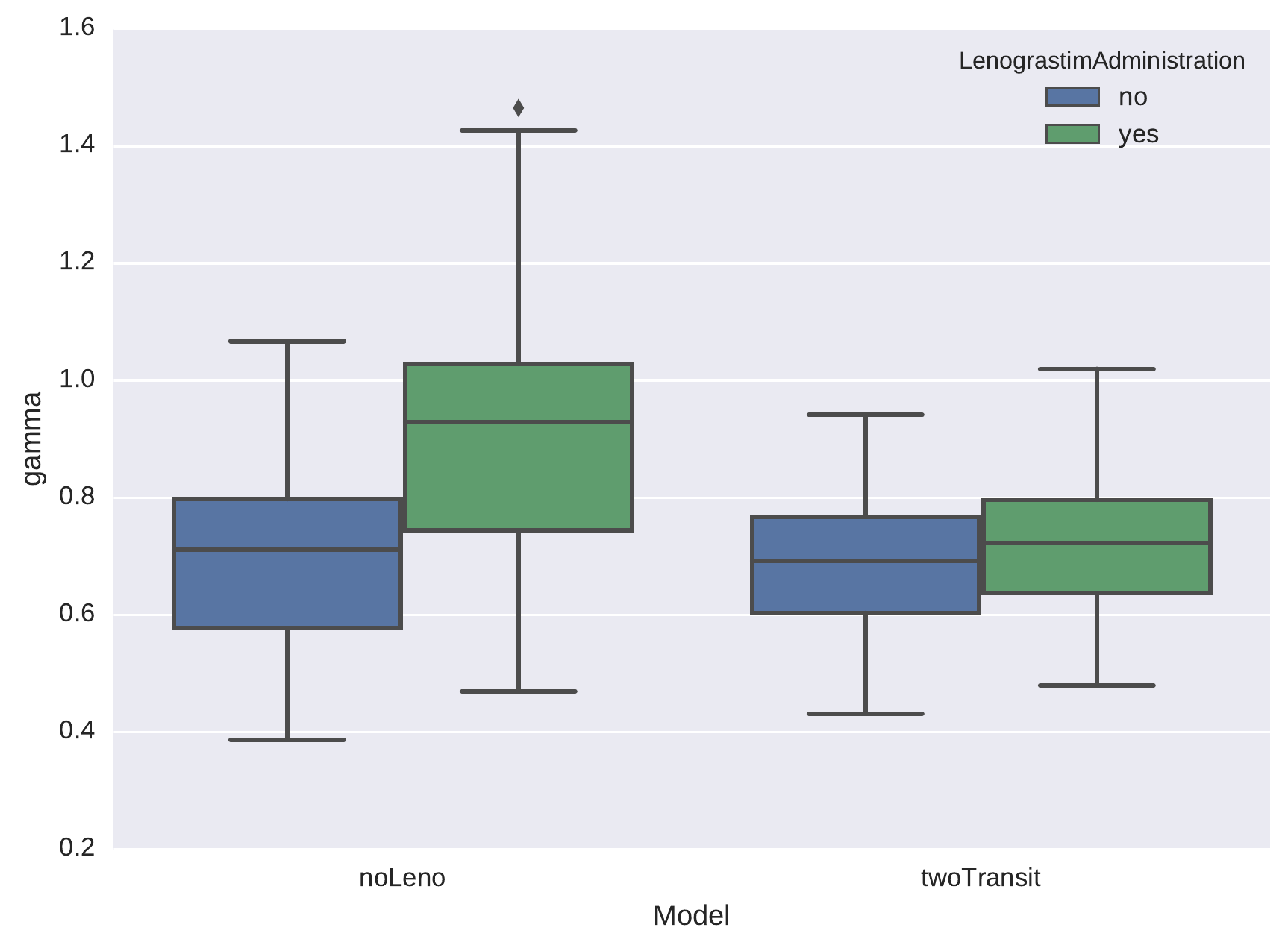} 
\end{center}
%\subcaption{}
\end{subfigure}
\begin{subfigure}[c]{0.49\textwidth}
\begin{center}
\includegraphics[width=1.\textwidth]{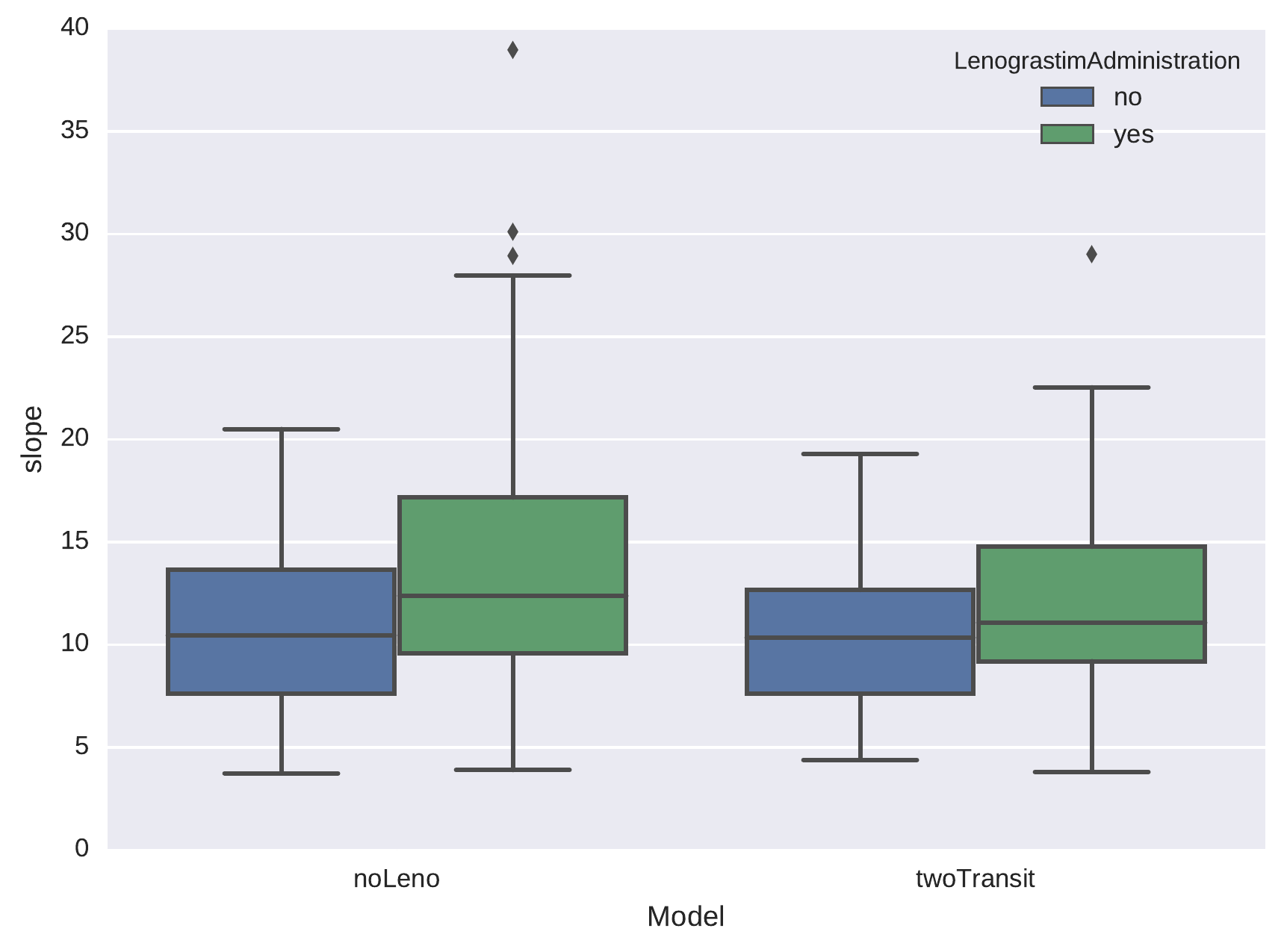} 
\end{center}
%\subcaption{}
\end{subfigure}
\begin{subfigure}[c]{0.49\textwidth}
\begin{center}
\includegraphics[width=1.\textwidth]{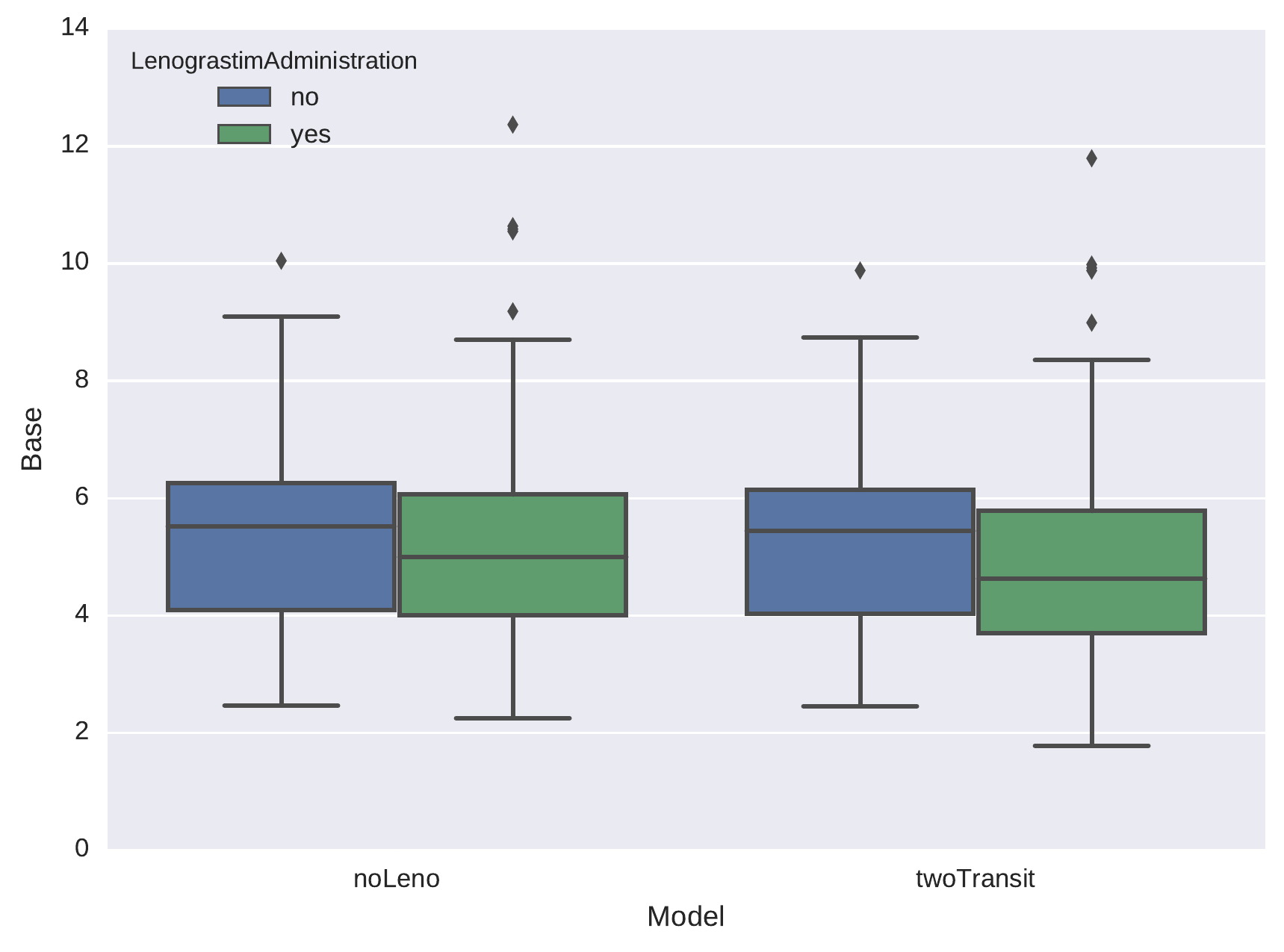} 
\end{center}
%\subcaption{}
\end{subfigure}
\caption{Boxplots of final parameter values of $\mktr, \mgamma, \mslope$ and $\mbase$ grouped by lenograstim administrations for models \textit{noLeno} and \textit{twoTransit} (see Table~\ref{tab:LenoModel}).}
\label{fig:boxplotsLenograstim}
\end{figure*}

\begin{figure*}[!t]
\centering
\includegraphics[width=0.9\textwidth]{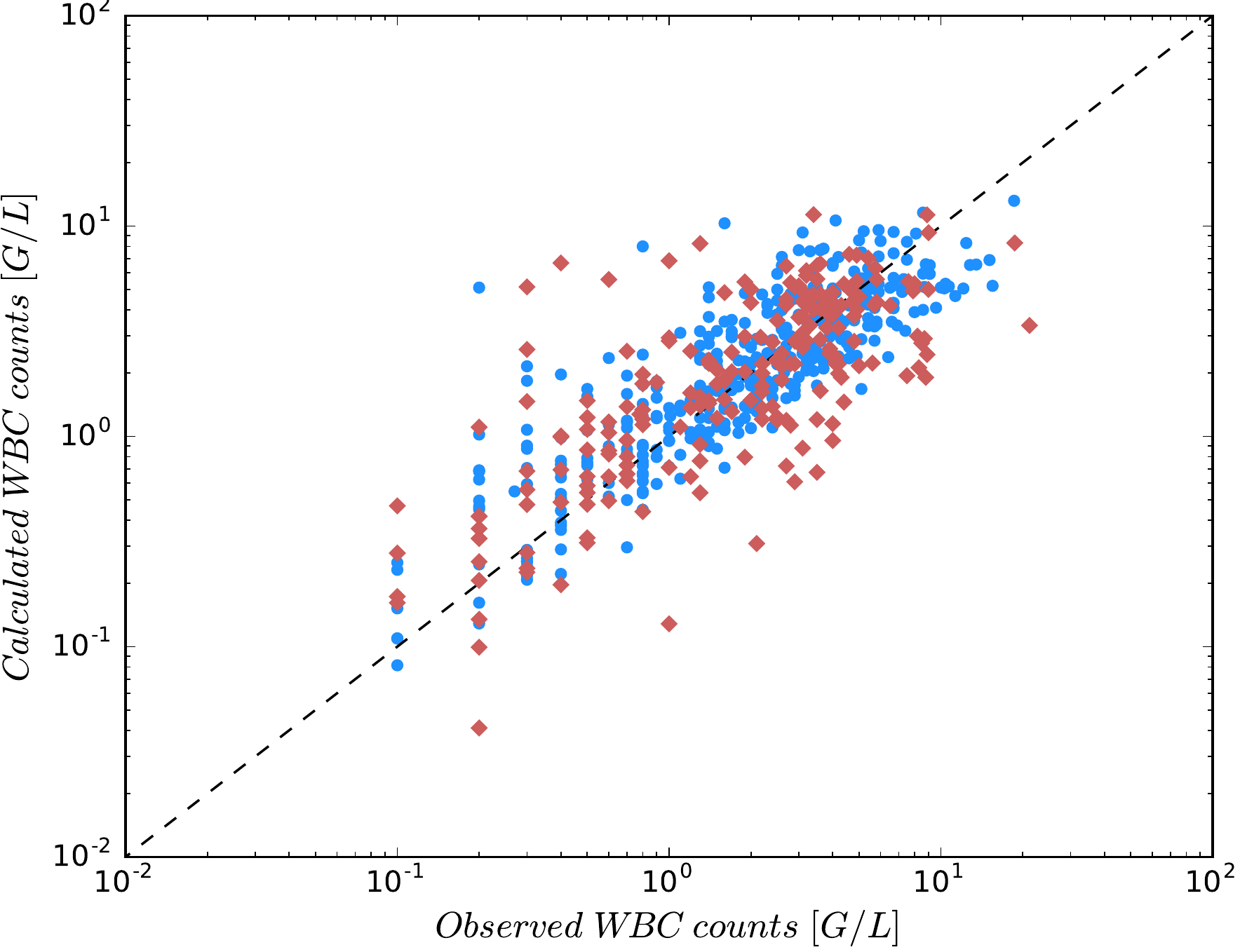} 
\caption{Goodness-of-fit plot visualizing observed versus individually calculated white blood cell (WBC) counts for 24 patients. 
%One patient was excluded as the model already converged to its leukemic steady state in the last consolidation cycle (CC).
Models were cross validated using out-of-sample (from the last CC) WBC measurements (blue circles).
Predictions were performed for the last cycles of the patients shown in red.
Red cubes show in-sample WBC measurements from the remaining first CCs.
}
\label{fig:Prediction}
\end{figure*}

\begin{figure*}[!t]
\begin{center}
\includegraphics[width=.99\textwidth]{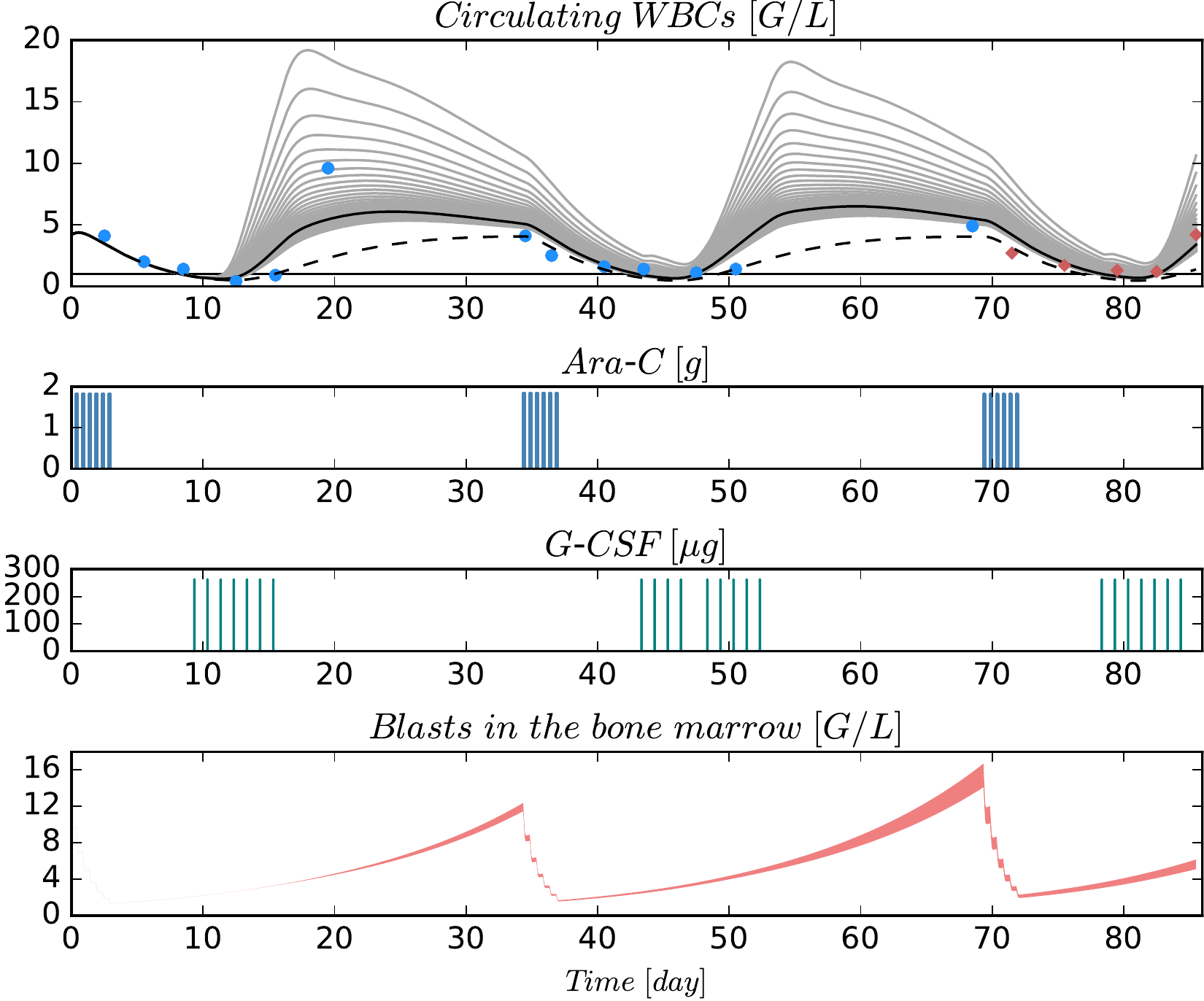} 
\end{center}
\caption{WBC dynamics (solid black line) of the final model fitted to observed WBC counts of the first two consolidation cycles (blue dots), and the Ara-C and lenograstim treatment schedules of one exemplary patient are shown.
The last cycle is used for model prediction and out-of-sample comparison. 
Simulated WBC dynamics for no lenograstim (dotted black line) and for different G-CSF steady state values (from $20\%$ to $140\%$ of used value) are shown (solid gray lines). 
No lenograstim administration prolonged WBC recovery time and lower/higher G-CSF steady state values shortened/prolonged WBC recovery.
Moreover, no lenograstim administration resulted in a slightly larger leukemic blast count (the red area indicates the difference when compared to the actual treatment schedule shown in the third row). 
}
\label{fig:GCSFsensitivity}
\end{figure*}

\subsection{Model development and fitting}

In a first step, the Ara-C version of Quartino's myelosuppression model was extended through absorption models with varying transit compartments decribing the s.c. administration of lenograstim \cite{Kagan2014} and a bioavailability of 30\% determined in \cite{Fernandez2007}.
The models were fitted to a variety of different consolidation cycles to determine the absorption model which described the hematopoietic effects of lenograstim administration best.
The final model contains two absorption rate constants $k_{a1}$ and  $k_{a2}$ and a bioavailability of 100\% similar to s.c. modeling approaches presented in \cite{Kagan2014}.
A bioavailability of 30\% and one absorption rate constant resulted in non-identifiability issues.
As no concentration measurements of G-CSF were collected and the global sensitivity analysis (compare for \ref{fig:sensitivityAnalysis}) revealed that the G-CSF related constants did not significantly contribute to the model outcomes, we moved forward with the best possible model.
As two distinct absorption rate constants are not identifiable \cite{Kagan2014}, we assumed a linear relation between $k_{a1}$ and  $k_{a2}$ and tested several factors until we defined the relation $k_{a2}=\frac{10}{3} k_{a1}$.

In a second step, a cytokine-dependent two-compartment model describing the dynamics of leukemic blasts was incorporated and the complete PK/PD model was fitted to the dataset of the Ara-C consolidation arm of the AMLSG 12-09 trial.
For the analysis of the influence of the leukemic blast lineage on the WBC lineage, we performed two parameter estimations with and without consideration of leukemic blasts.

\subsection{Summary of~\ref{fig:VPC_GoF}}
\ref{fig:VPC_GoF} shows visual predictive checks for the circulating WBCs and relative blasts counts in the BM derived from the final model.
The median of observed WBC counts coincided with the median of calculated WBC counts and fell within or close to its $95\%$ prediction interval (blue area).
The 2.5th percentile of the model shows an underestimation in the first CC and the 97.5th model percentile shows overestimations in all CCs.
Considering the VPC of the blasts, the $95\%$ prediction intervals of the 50th and 97.5th percentiles indicated that the model assumed a faster increase of blasts during the three CCs compared to the almost constant (2.5th and 50th percentiles) and decreasing (97.5th percentile) dynamics observed within the patients.

\subsection{Summary of~\ref{fig:boxplotsLenograstim}}
The boxplots of the four parameters in~\ref{fig:boxplotsLenograstim} in the SM highlight that $\mktr$ and $\mgamma$ were significantly increased in the model without consideration of an absorption model for patients who received lenograstim administrations.
After extending the model with the s.c. absorption model and two transit compartments the parameter values of $\mgamma$ were almost equal between the two groups whereas the $\mktr$ values only approached to a small degree.
Values of $\mslope$ were higher and the WBC steady state values were slightly lower in the group of lenograstim administrations.

\subsection{Summary of~\ref{fig:GCSFsensitivity}}
\ref{fig:GCSFsensitivity} shows the influence of G-CSF administrations (yes or no) and of varied G-CSF steady states on the WBC recovery.
Ara-C without lenograstim administration resulted in a longer recovery time and a slightly lower WBC count before the start of the second and third CC.
As a further consequence, the number of leukemic blasts in the BM was higher and increased more over time, than to the leukemic blast count when the actual treatment schedule of lenograstim was conducted. 
A different G-CSF steady state affected the recovery time, where lower steady-state values provoke an overproduction of WBC counts, leading to a higher value than the WBC steady state.

\subsection{Summary of~\ref{fig:sup_OCsolutionOne} to \ref{fig:sup_OCsolutionSeven}}
\ref{fig:sup_OCsolutionOne} to \ref{fig:sup_OCsolutionSeven} show detailed optimization results for 14 patients from the data set (those with at least two consolidation cycles and relative bone marrow blast count measurements). 
The plots are identical to those in Figure~\ref{fig:OCPsolution} from the manuscript but with an additional row for the dynamics of endogenous G-CSF, and the number of patients from 1 to 14 is identical to that in Figure~\ref{fig:OptimizedNadirsBlasts}. Optimal solution refers to the objective function, the model, and constraints as specified in the manuscript.

The treatment schedules of Ara-C and lenograstim in the last CC were optimized (black) and compared with the clinically applied treatment schedules (red). 
WBC counts and relative blast counts in the bone marrow (BM) resulting from individual models (blue dotted lines) are shown. 
Personalization was performed using in-sample measurements (blue dots) and clinical treatment schedules (blue lines) from all but the last CC.

While optimal timing and dosages of Ara-C and lenograstim were personalized and hence different for each considered patient, two qualitative patterns are observed. 

In pattern A, an additional Ara-C administration period (and hence an additional CC) is introduced; the administration order is Ara-C, lenograstim, Ara-C and lenograstim.

In pattern B, the nadir is increased compared to the clinical treatment schedule with the administration order lenograstim, Ara-C, lenograstim. The amount of Ara-C is usually considerably reduced.

The pattern types (9 times A, 5 times B) are indicated in the captions for convenience.

\subsection*{Sensitivity analysis}

A global sensitivity analysis was conducted to identify the impact of each parameter, respectively constant, on the variability of two model outputs \cite{Zhang2015}.
The model outputs of interest were the leukemic cells in the bone marrow ($x_{l1}$) at the end of a consolidation cycle and the nadir of circulating white blood cells similar to the two objective function terms in the optimization problem \eqref{eq:OCProblem}.
The sensitivity analysis was performed in R (version 3.6.1) using the packages \textit{mrgsolve} (version 0.10.0) for solving the ODE system \eqref{eq:ODEmodel} and \textit{sensitivity} (version 1.17.0, \textit{sobolmartinez} function) for the global sensitivity analysis.
The function \textit{sobolmartinez} implements the Monte Carlo estimation of the Sobol’ indices for both first-order and total indices for each parameter using correlation coefficient-based formulas. 
These are called the Martinez estimators.
The Sobol method is based on the decomposition of the model output variance into fractional contributions from effects of single parameters considering no interaction between parameters (first-order/main effect) or an interaction between two (second-order), more or all (total effect) parameters.
%The sum of Sobol’ indices is 1 such that to each parameter its relative influence on the model output can be assigned.
%Sobol method is based on the decomposition of the model output variance D into contributions from effects of single parameters, combined effects of pairs of parameters, and so on.
%first-order indices no consideration of interaction between parameters
%total order quantifies the overall effects of one parameter on the model output.
%main effect (first-order) are used to measure the fractional contribution of a single parameter to the output variance.
The theoretical background can be found in the tutorial \cite{Zhang2015} and references therein.
The experimental setup was chosen as follows.
One consolidation cycle was defined with cytarabine schedule D123 for an exemplary patient with a BSA of $1.8 \ m^2$ and six 263 $\mu g$ daily s.c. lenograstim administrations starting at day 9. 
Two datasets with 3800 (2 times 19 [number of variables] times 100) sets of parameters were generated with uniformly chosen parameter samples from a 0.5-fold decrease to a 1.5 fold increase in the nominal parameter values.
%Explain how dataset was generated. plus minus 40\%
%100 uniformly chosen samples from a 0.6-fold decrease to a 1.4 fold increase in the nominal parameter values.
%The function generates uniform samples from a 100 fold decrease to 100 fold increase in the nominal parameter value.
%%The return value is a list with two data frames that can be passed into the Sobol function.
%One cycle with cytarabine administration  D123 and 6 263 $\mu g$ daily s.c. lenograstim administrations starting at day 9. 
The contribution of each parameter to the variability in the two model outputs is presented in~\ref{fig:sensitivityAnalysis}.
The leukemic cells in the bone marrow as well as the nadir of circulating WBCs were influenced by the elimination constant $k_{10}$, the volume of the central compartment $V$ and the pharmacodynamic effect $\mslope$.
The leukemic cells in the bone marrow are further influenced by the fraction constant $a_1$ determining the fraction of daughter cells staying at the current differentiation stage and the stem cell proliferation $p_1$.
The nadir of circulating WBCs is additionally affected by $\mbase \ (\text{Base}), \mktr, \mgamma$ and $\mkma$.
The global sensitivity analysis reveals that the WBC nadir is mostly influenced by the hematopoietic parameters which we individually determined during parameter estimation and therefore we achieved good accuracies in predicting the nadir of the last CCs (c.f. \ref{fig:Prediction}).
We determined the dynamics of the leukemic cells in the bone marrow and circulating blood via the estimation of the initial value of the leukemic cell count in the bone marrow.
The sensitivity analysis showed that $a_1$ and $p_1$ had a larger impact compared to the initial value but the current data availability did not allow to estimate those parameters such that we fixed them to published values.


\figuresAppendix{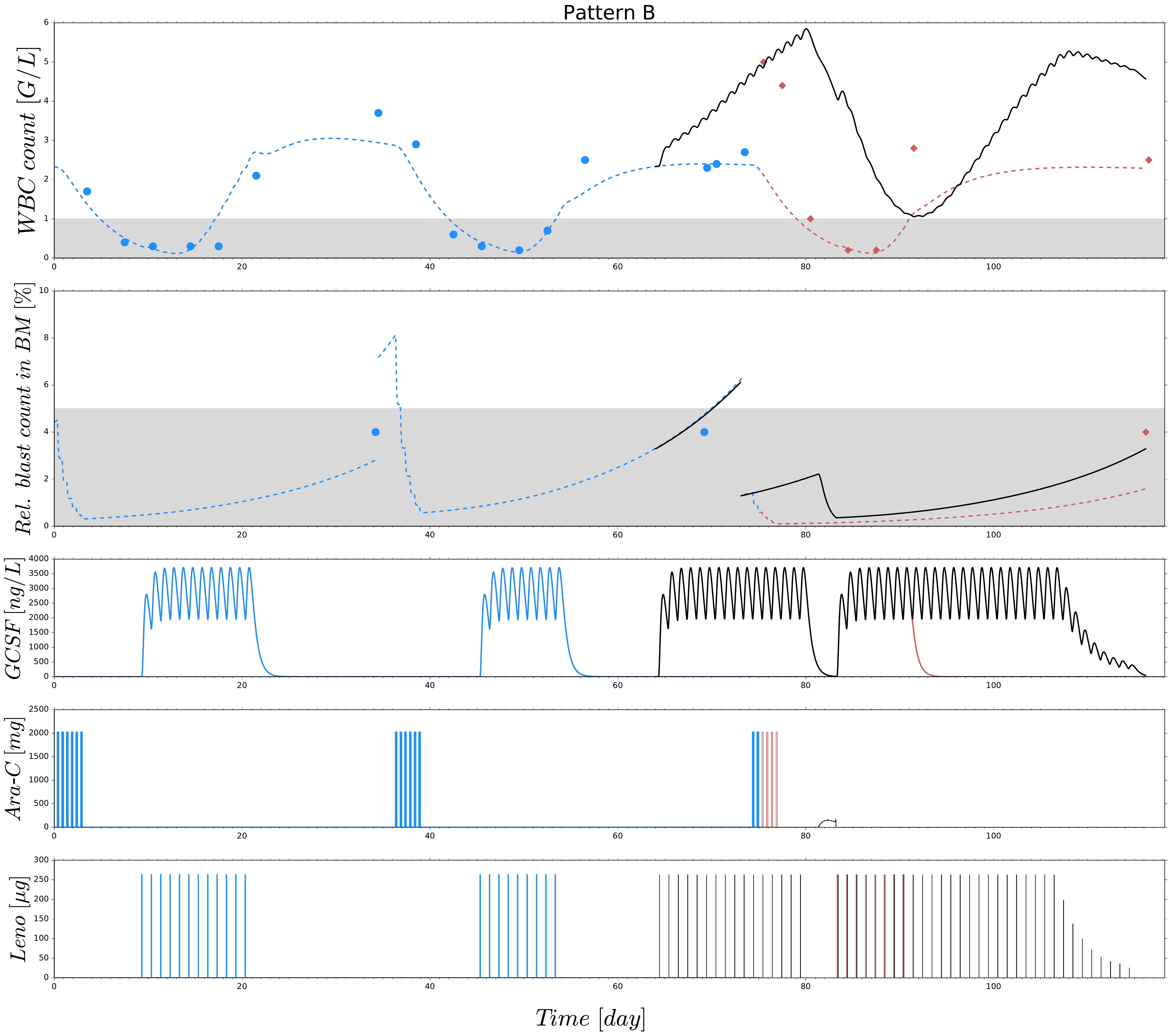}{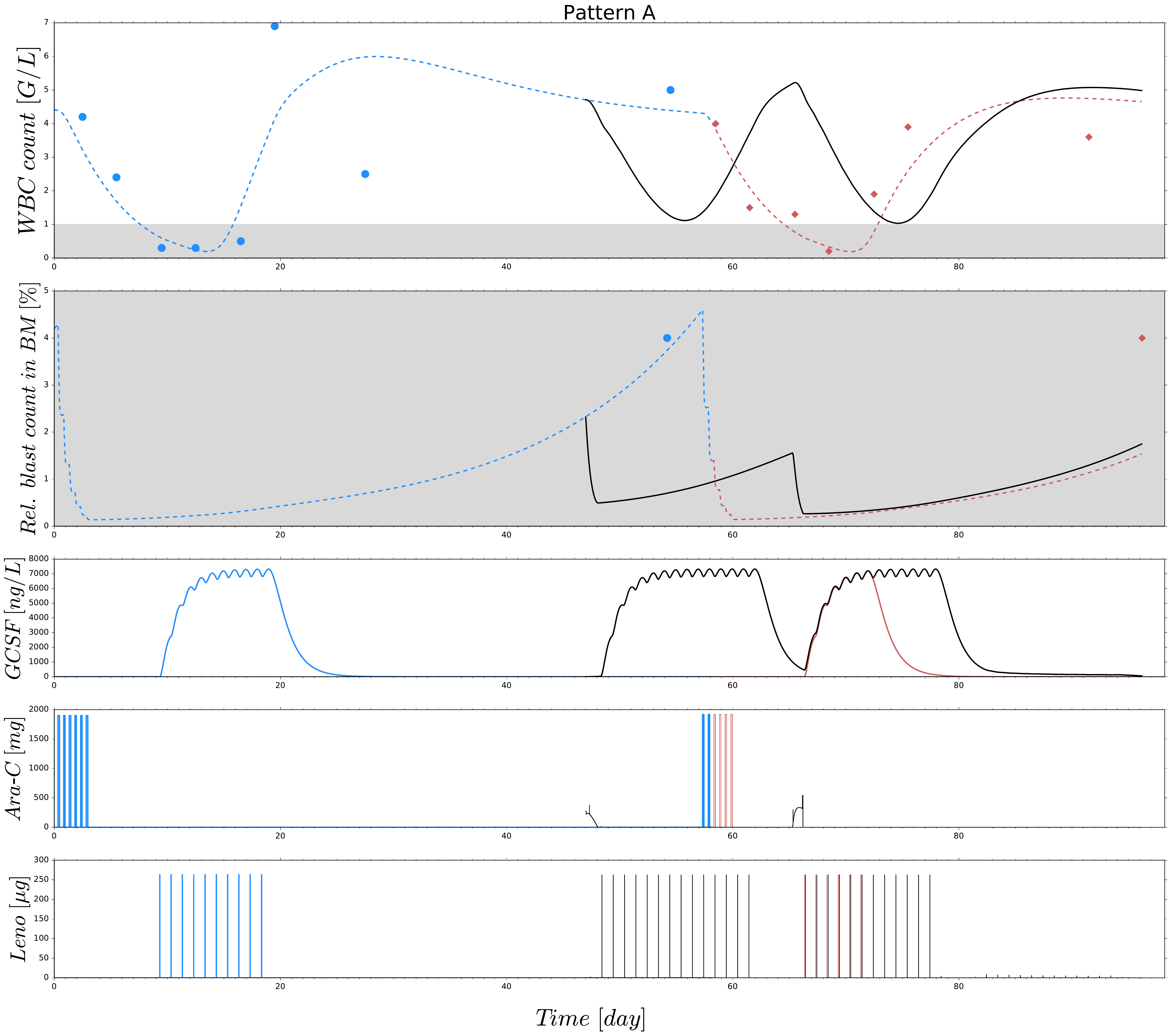}{fig:sup_OCsolutionOne}{1}{2}{B}{A}
\figuresAppendix{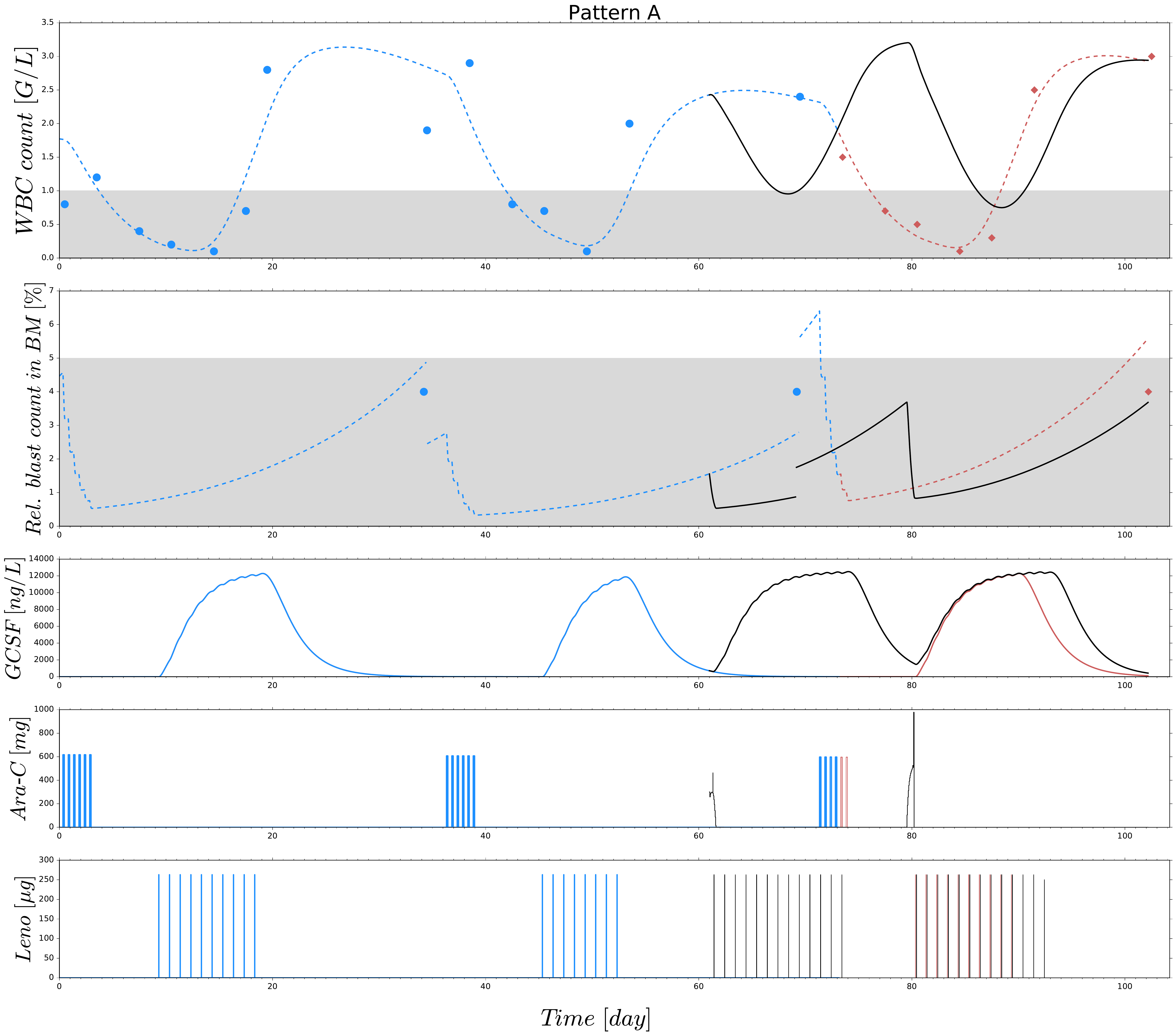}{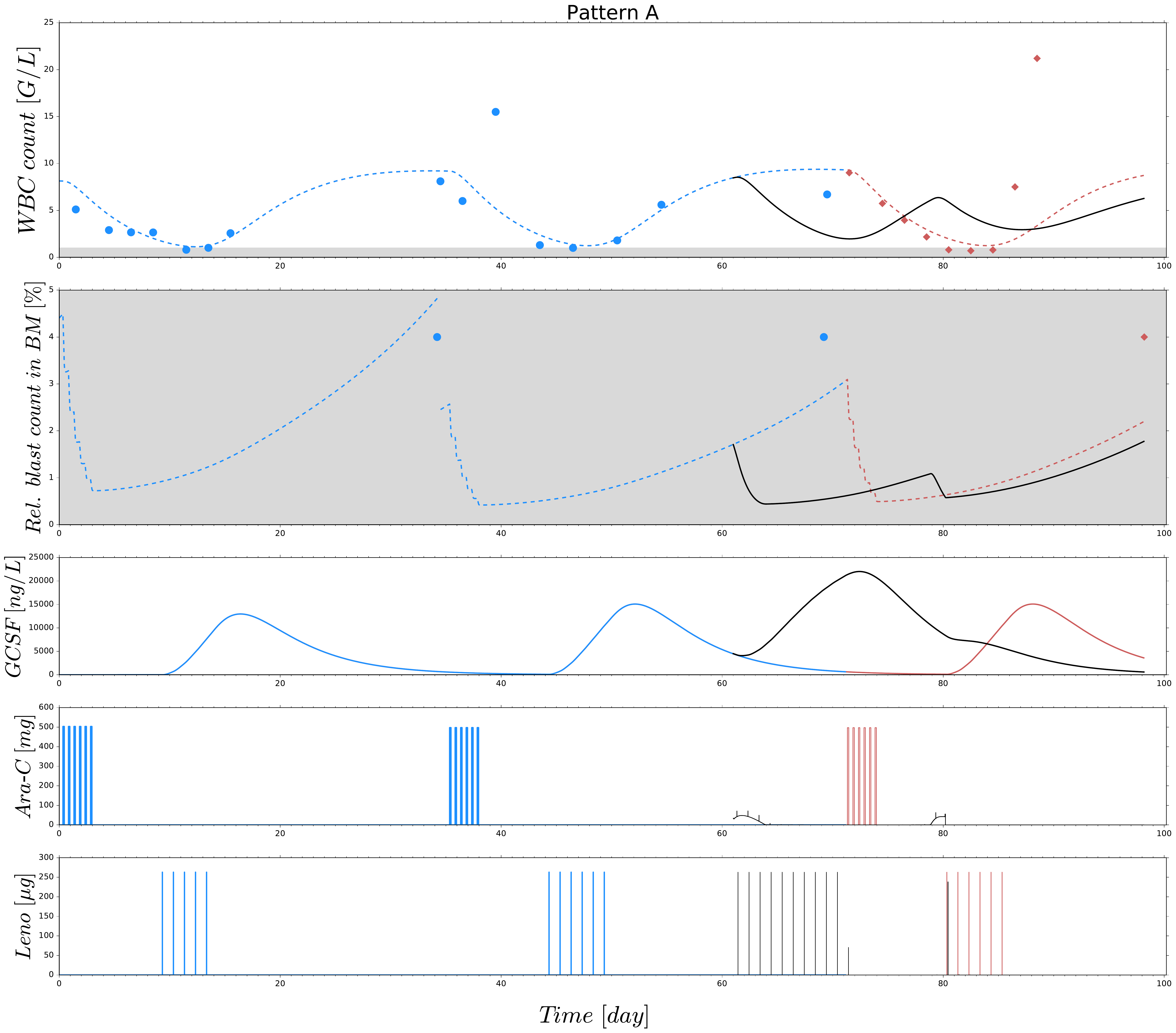}{fig:sup_OCsolutionTwo}{3}{4}{A}{A}
\figuresAppendix{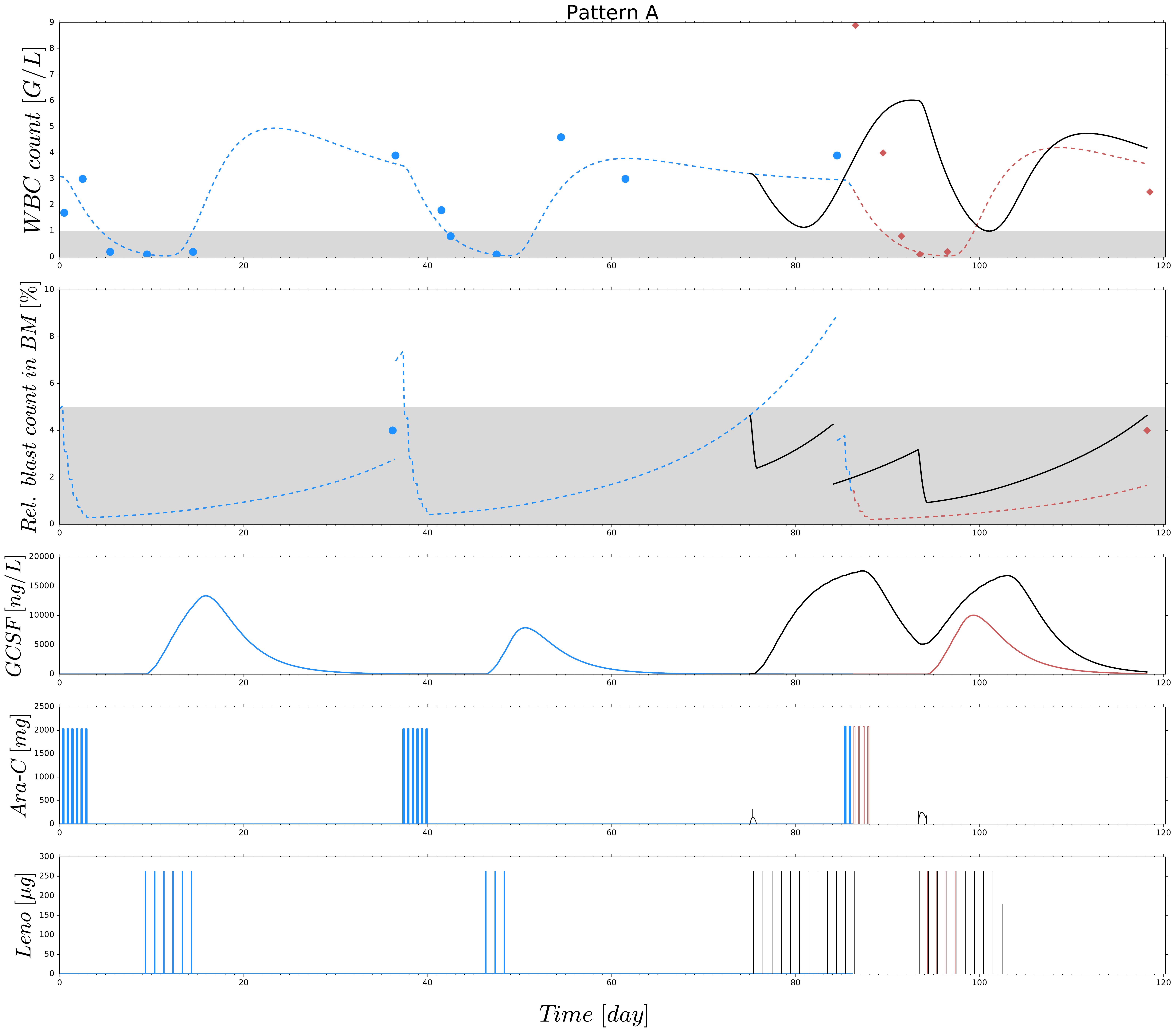}{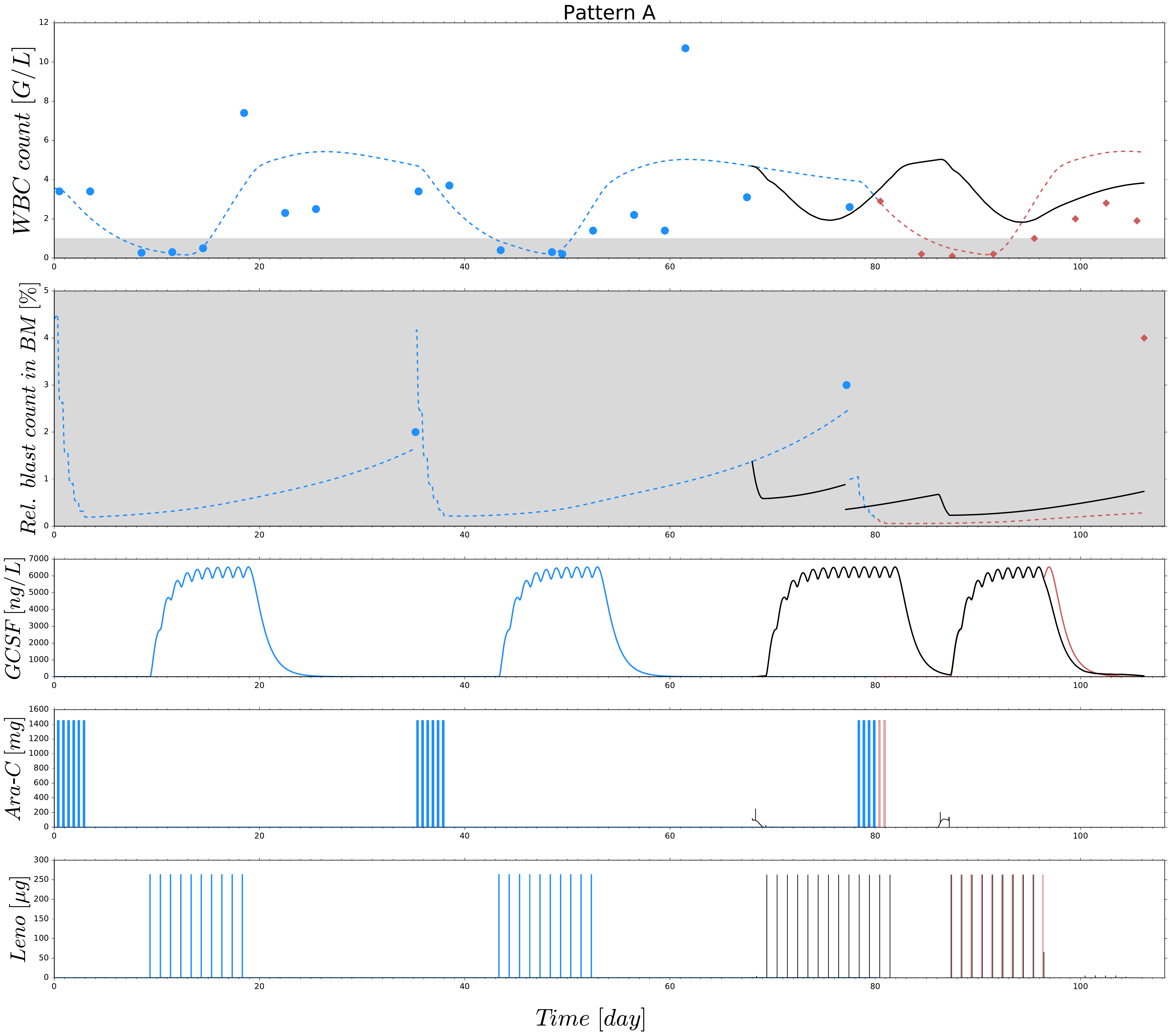}{fig:sup_OCsolutionThree}{5}{6}{A}{A}
\figuresAppendix{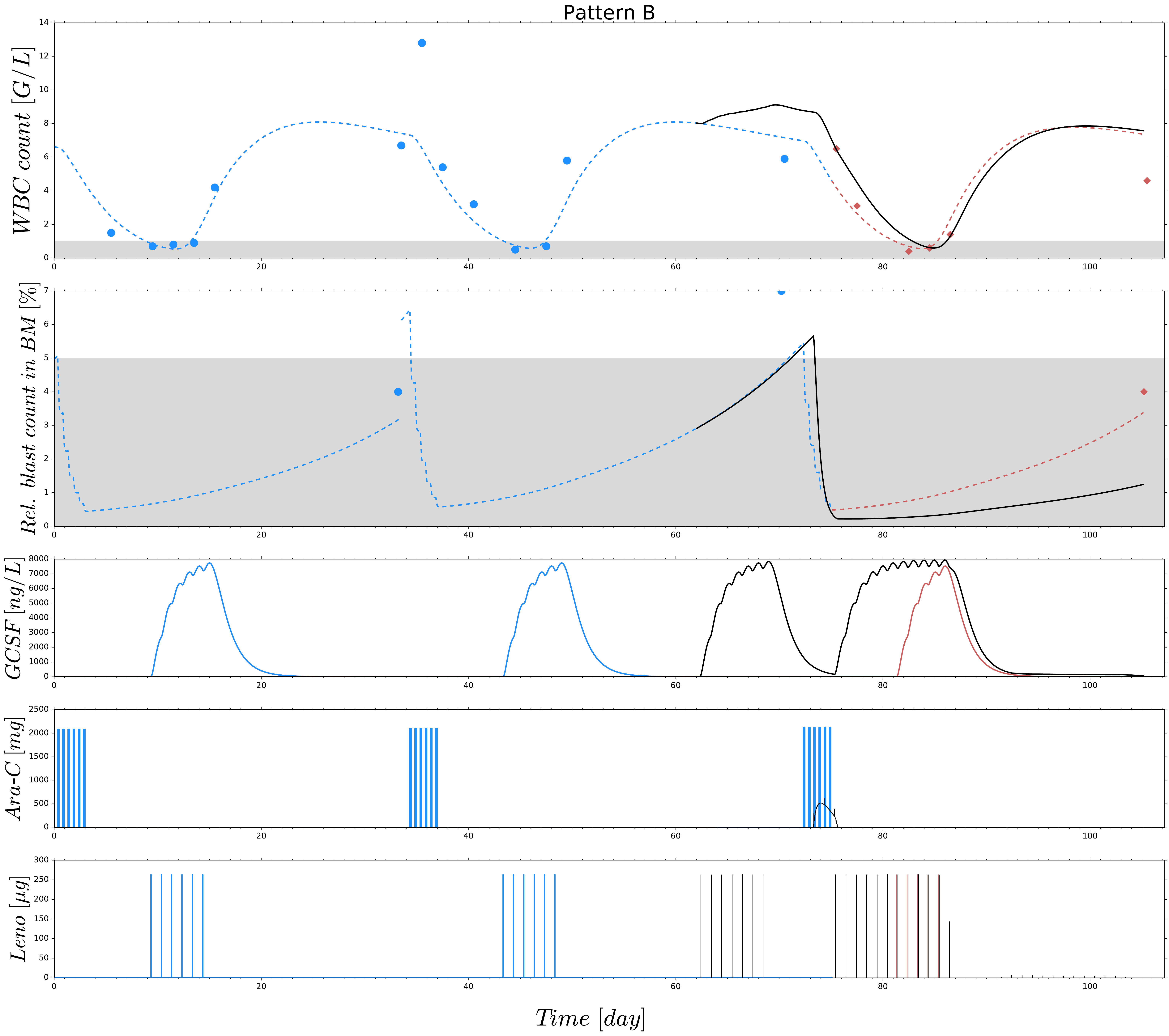}{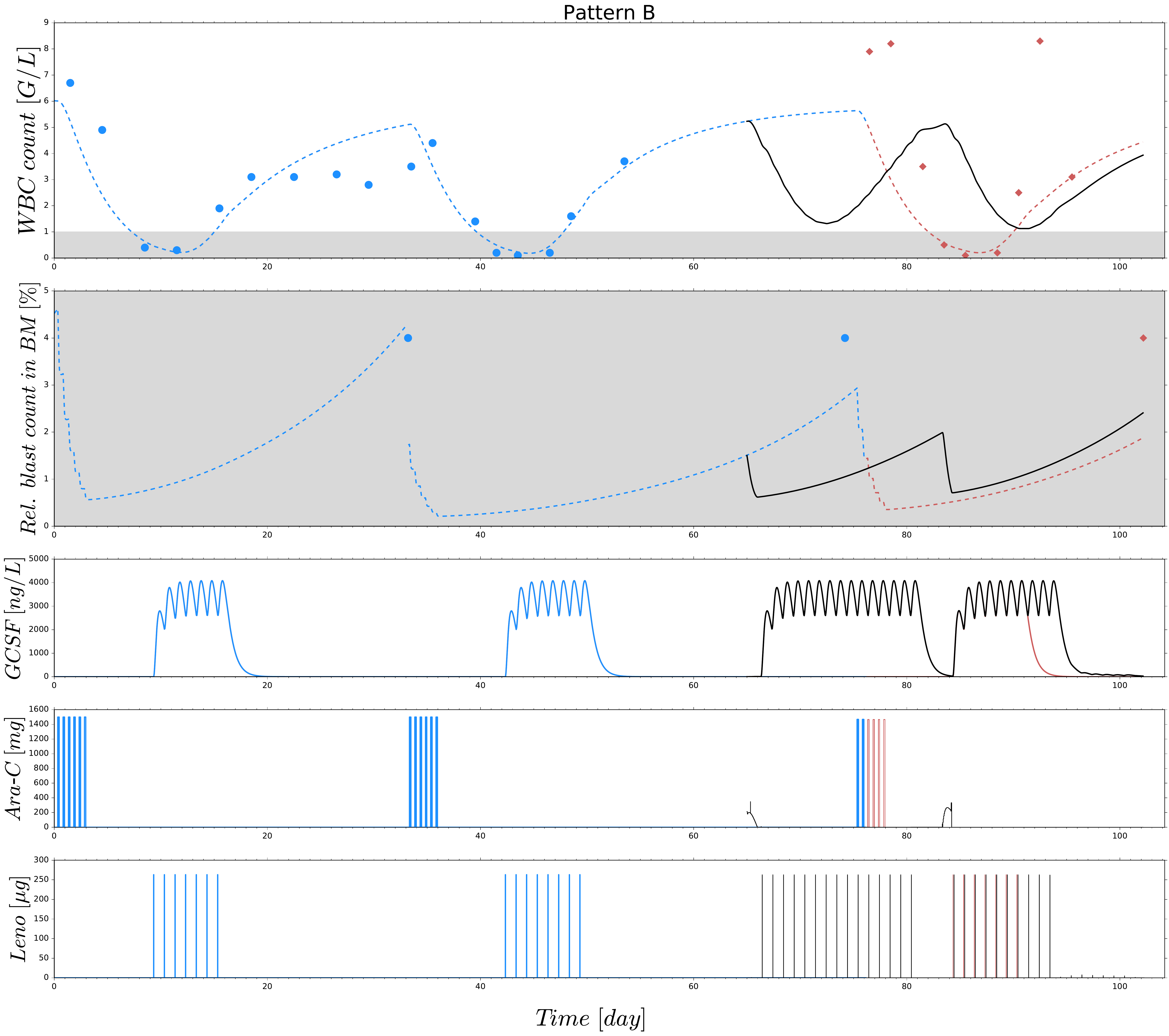}{fig:sup_OCsolutionFour}{7}{8}{B}{B}
\figuresAppendix{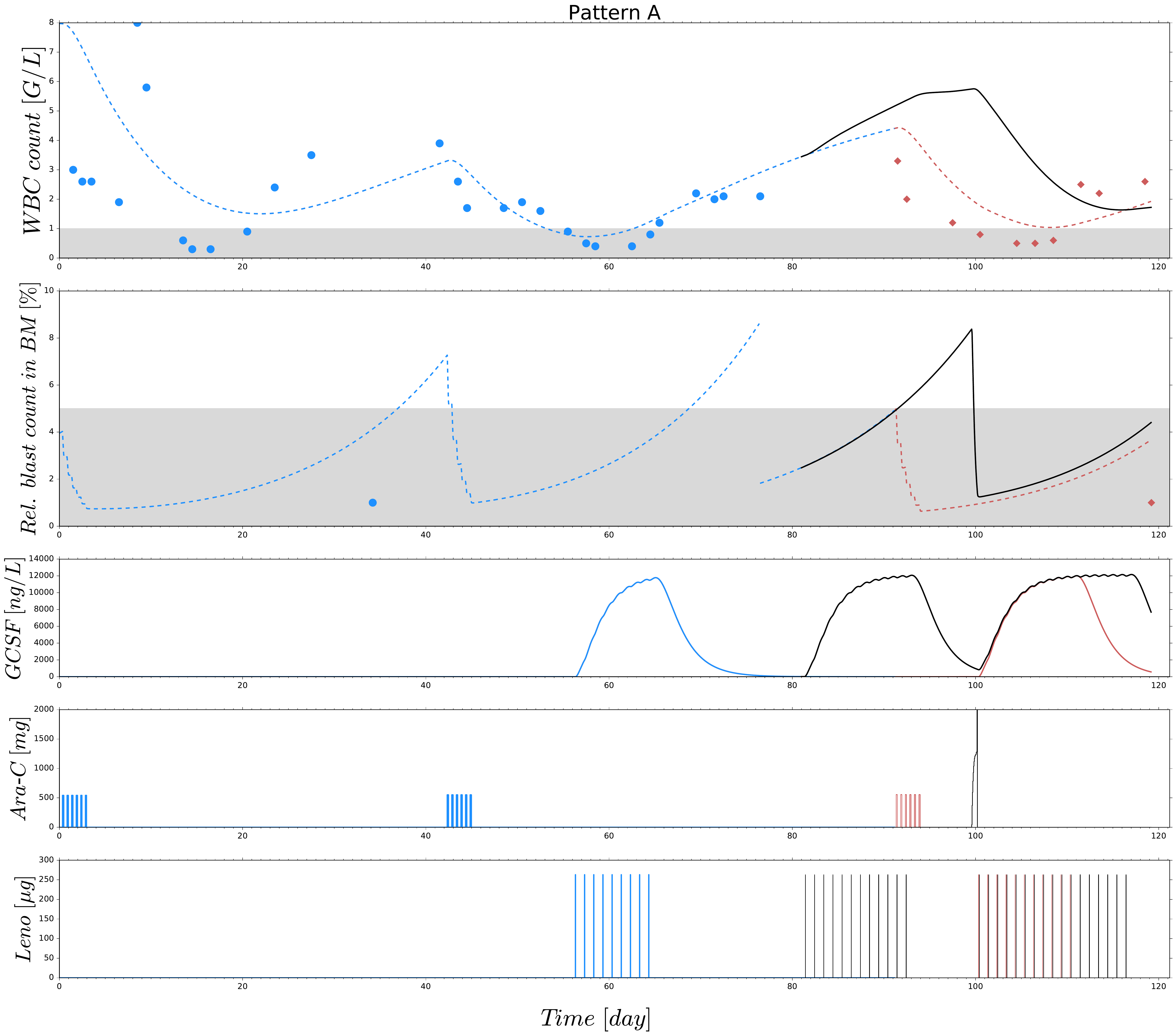}{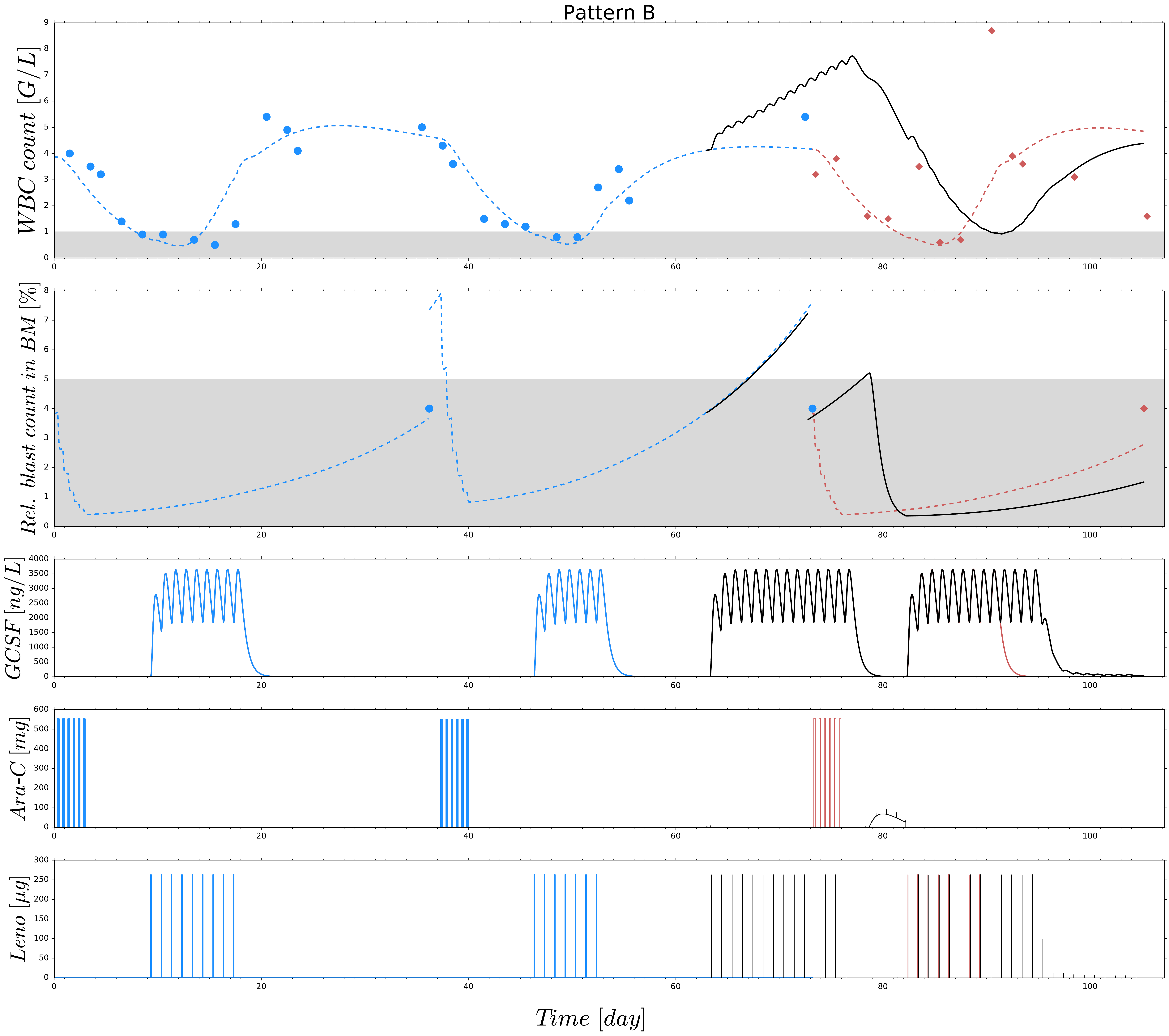}{fig:sup_OCsolutionFive}{9}{10}{A}{B}
\figuresAppendix{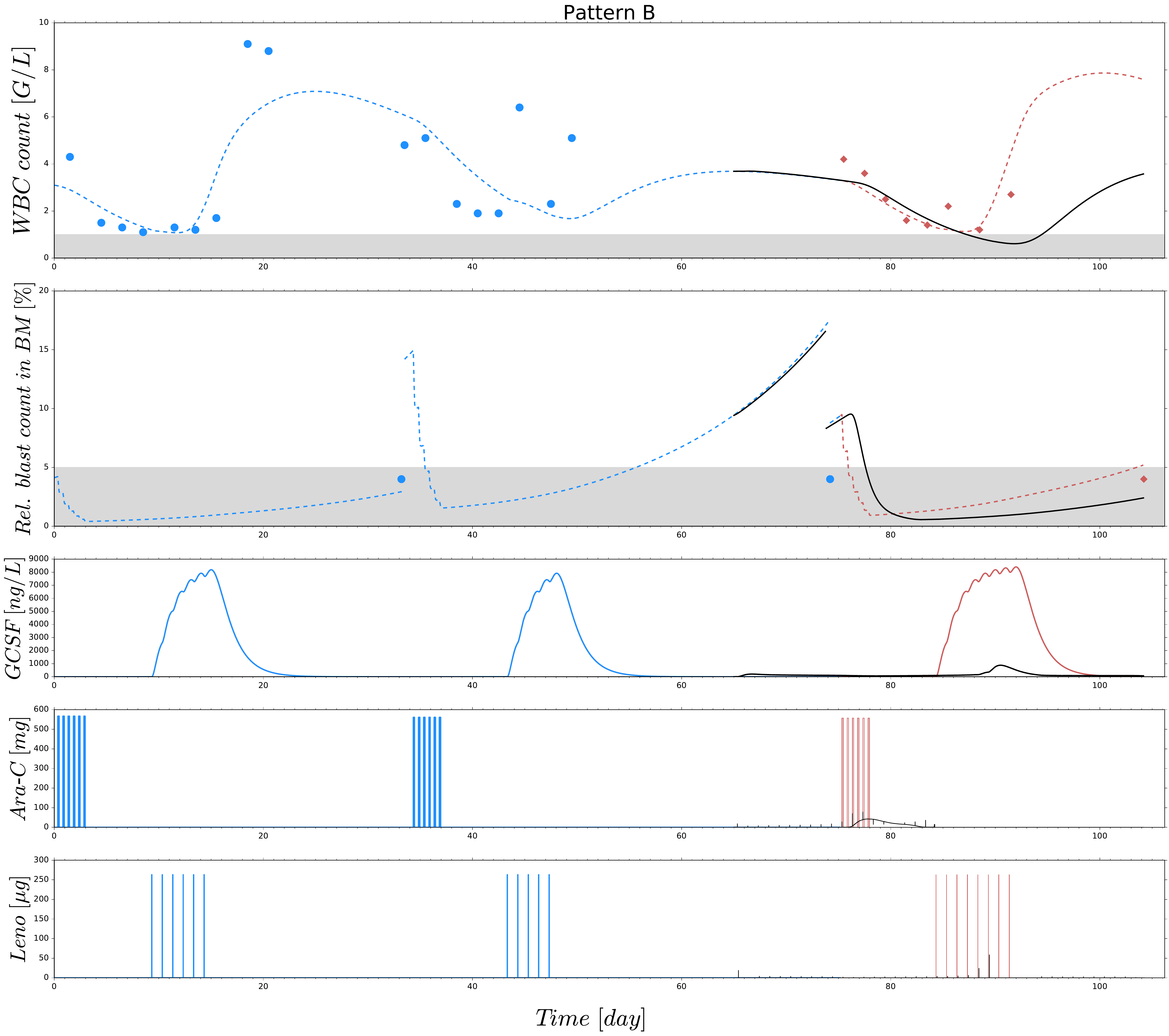}{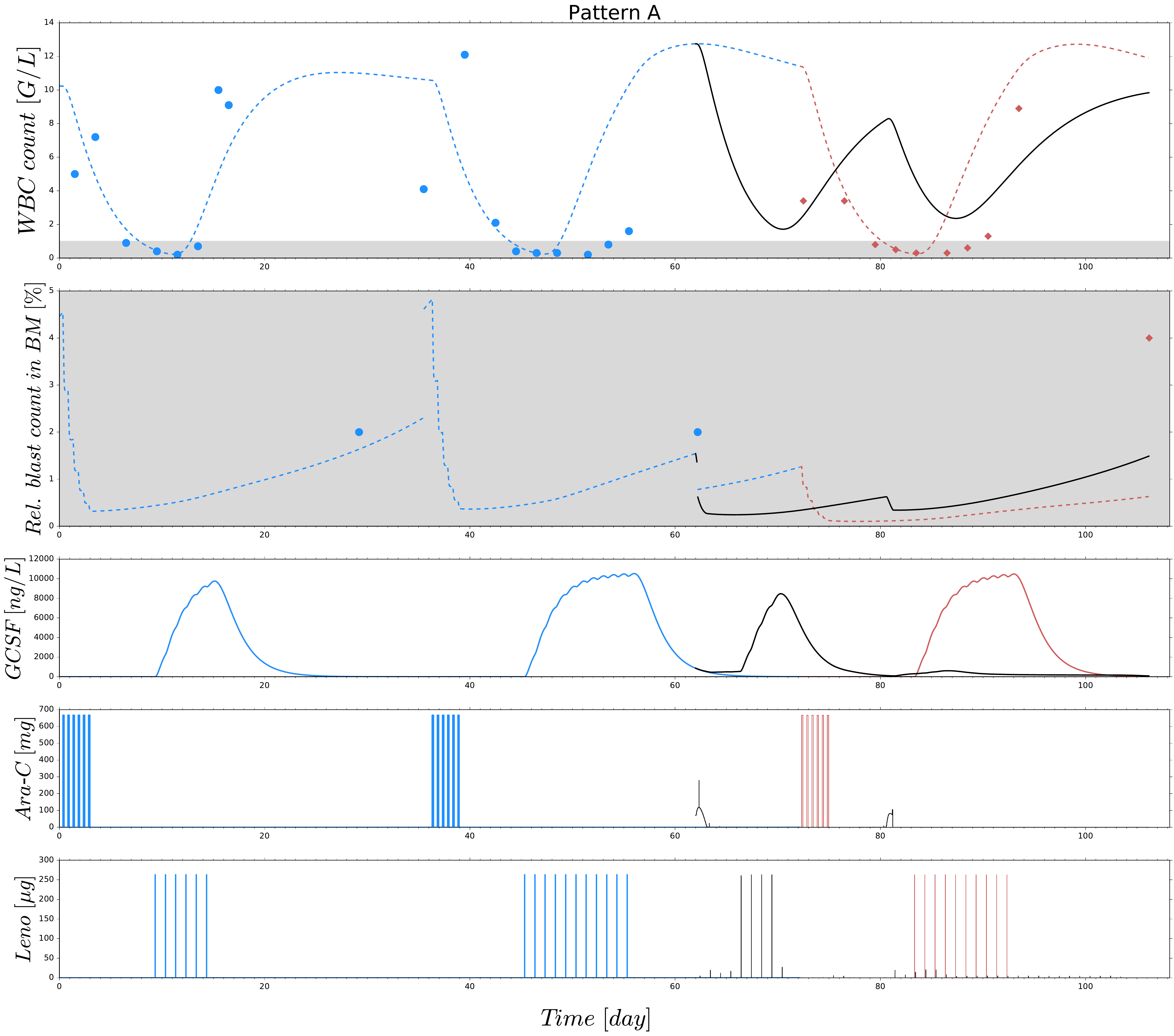}{fig:sup_OCsolutionSix}{11}{12}{B}{A}
\figuresAppendix{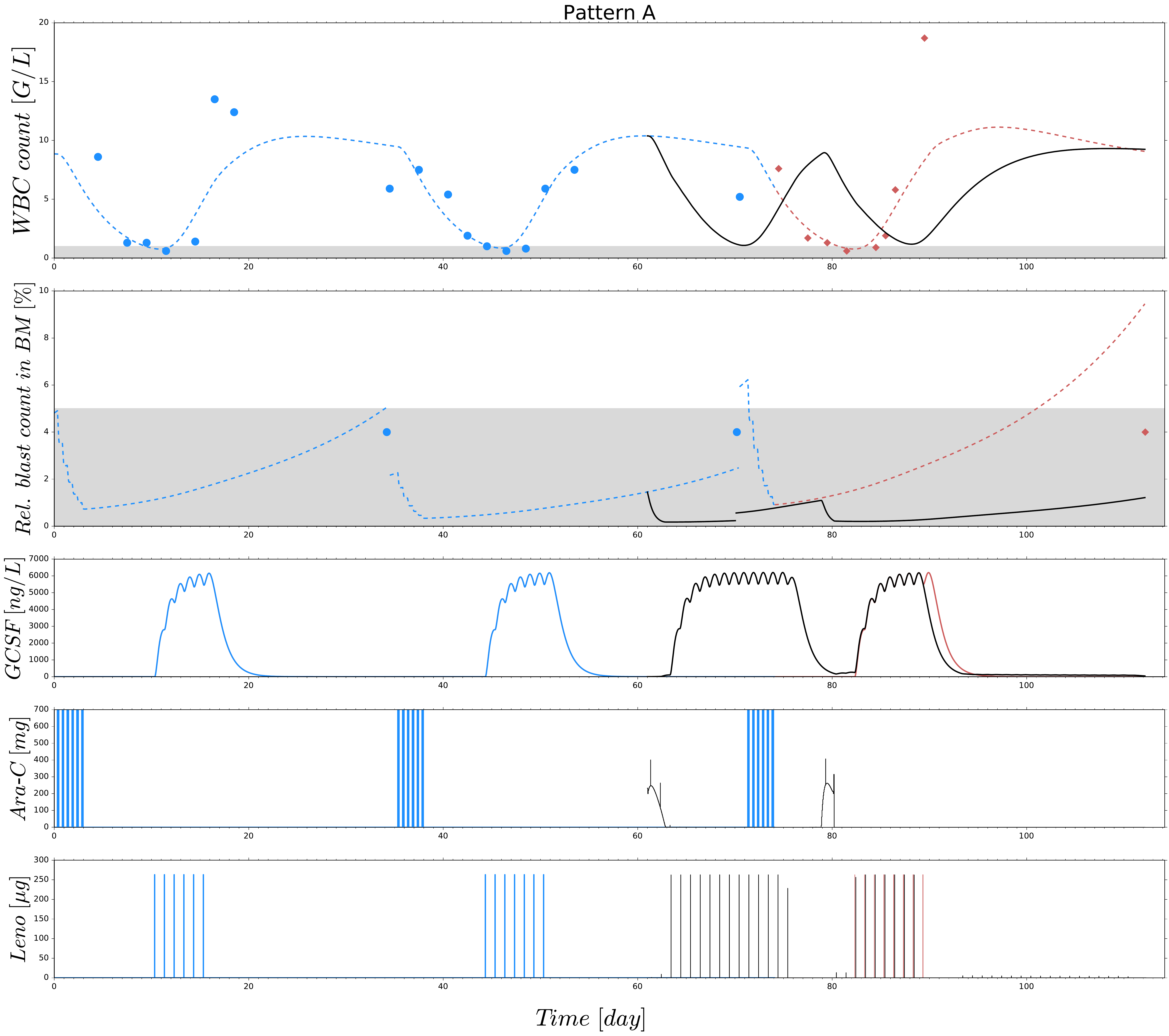}{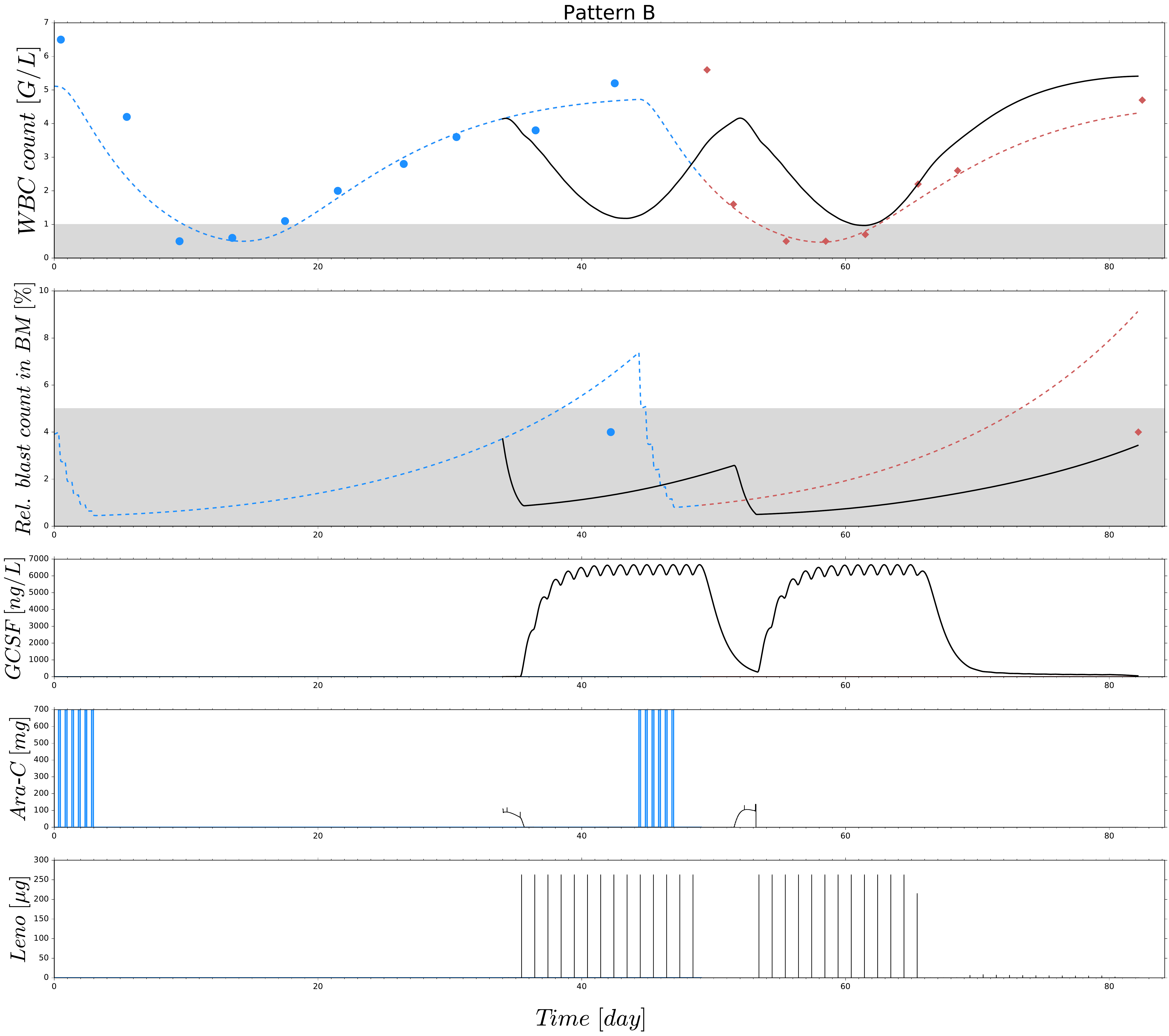}{fig:sup_OCsolutionSeven}{13}{14}{A}{B}

%myPattern = ["A","A", # 103 109
%             "B","A", # 115 118
%             "A","B", # 119 121
%             "A","A", # 123 126
%             "A","B", # 129 136
%             "B","A", # 138 141
%             "A","B"] # 143 144 

\begin{figure*}[!t]
\begin{subfigure}[c]{0.89\textwidth}
\begin{center}
\includegraphics[width=1.\textwidth]{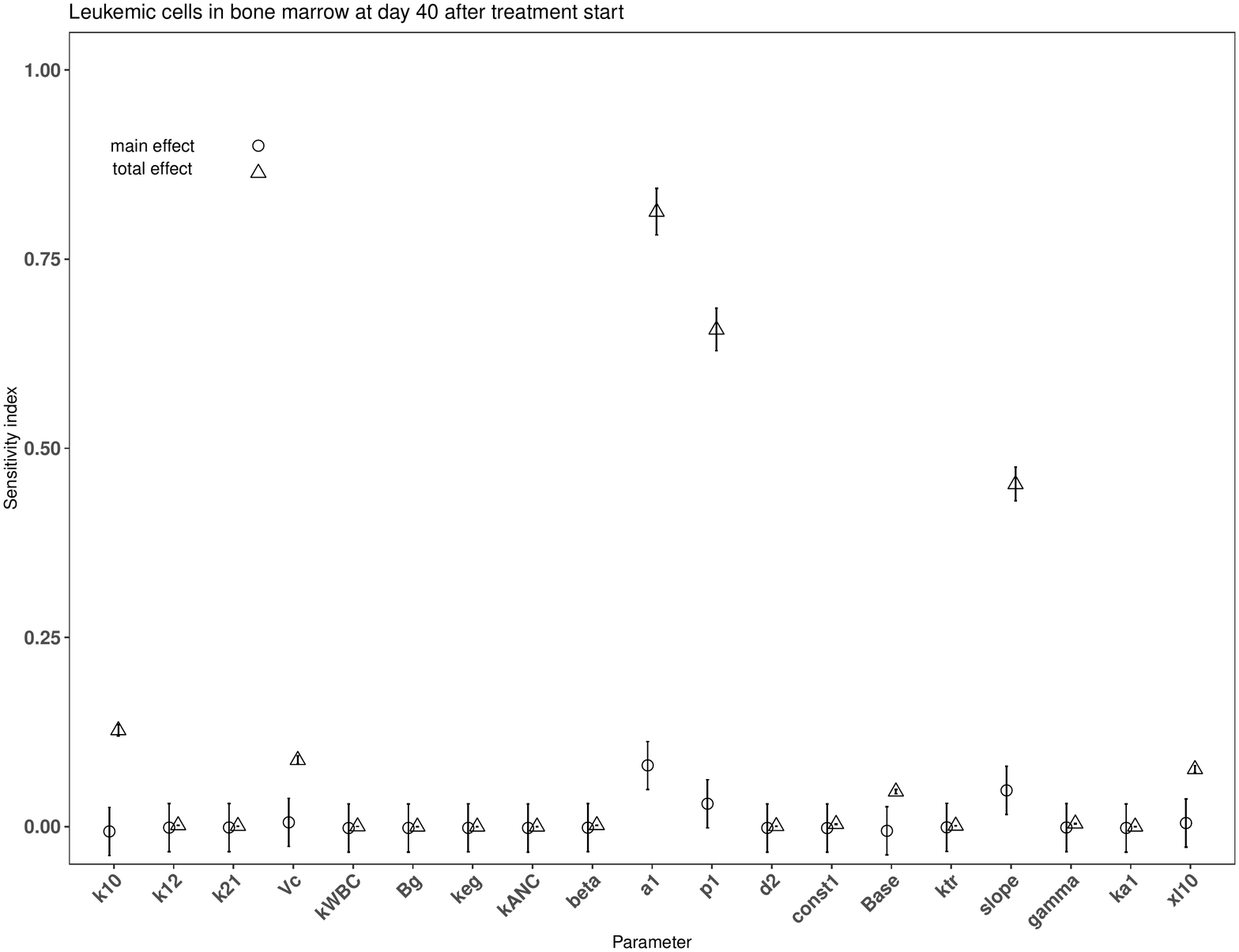} 
%\subcaption{}
\end{center}
\end{subfigure} 
\begin{subfigure}[c]{0.89\textwidth}
\begin{center}
\includegraphics[width=1.\textwidth]{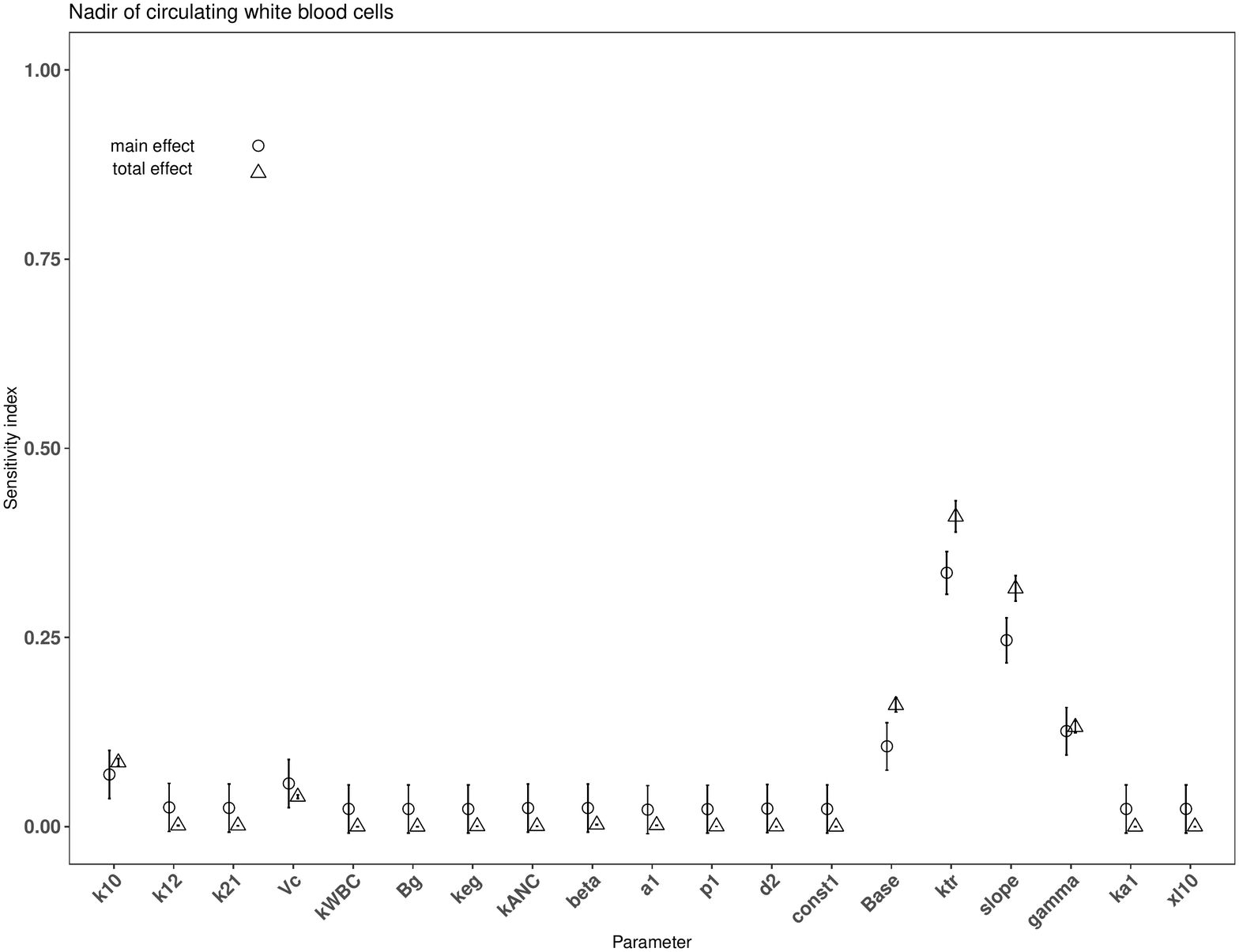} 
\end{center}
%\subcaption{}
\end{subfigure}
\caption{Global sensitivity analysis for the constants and parameters listed in \ref{tab:ModelSpecifications}.
Model outputs of interest were the amount of leukemic cells in the bone marrow at day 40 after treatment start and the nadir of circulating white blood cells.}
\label{fig:sensitivityAnalysis}
\end{figure*}

\ifCLASSOPTIONcaptionsoff
  \newpage
\fi

\end{document}